\documentclass[11pt,journal, onecolumn, draftcls]{IEEEtran}
\usepackage{amsmath}
\usepackage{epsfig}
\usepackage{amssymb}
\usepackage{graphicx}
\usepackage{hyperref}
\usepackage{color}
\usepackage{subfigure}
\usepackage{algorithmic}
\usepackage{algorithm}
\usepackage{psfrag}
\begin{document}
\title{Robust Adaptive Beamforming for General-Rank Signal Model
with Positive Semi-Definite Constraint via POTDC}
%\ninept
\author{Arash~Khabbazibasmenj,~\IEEEmembership{Student Member,~IEEE},
Sergiy A.~Vorobyov\thanks{S.~A.~Vorobyov is the corresponding
author. A.~Khabbazibasmenj and S.~A.~Vorobyov are with the
Department of Electrical and Computer Engineering, University of
Alberta, Edmonton, AB, T6G~2V4 Canada; e-mail: \{{\tt khabbazi,
svorobyo\}@ualberta.ca}. This work is supported in parts by the
Natural Science and Engineering Research Council (NSERC) of
Canada. Parts of this paper have been presented at CAMSAP~2011,
San Juan, Puerto Rico. },~\IEEEmembership{Senior Member,~IEEE}}
%
%\vspace{-1cm}
%
\maketitle
%
%\vspace{-1.5cm}
\begin{abstract}
The robust adaptive beamforming (RAB) problem for general-rank
signal model with an additional positive semi-definite constraint
is considered. Using the principle of the worst-case performance
optimization, such RAB problem leads to a difference-of-convex
functions  (DC) optimization problem. The existing approaches for
solving the resulted non-convex DC problem are based on
approximations and find only suboptimal solutions. Here we solve
the non-convex DC problem rigorously and give arguments suggesting
that the solution is globally optimal. Particularly, we rewrite
the problem as the minimization of a one-dimensional optimal value
function whose corresponding optimization problem is non-convex.
Then, the optimal value function is replaced with another
equivalent one, for which the corresponding optimization problem
is convex. The new one-dimensional optimal value function is
minimized iteratively via polynomial time DC (POTDC) algorithm. We
show that our solution satisfies the Karush-Kuhn-Tucker (KKT)
optimality conditions and there is a strong evidence that such
solution is also globally optimal. Towards this conclusion, we
conjecture that the new optimal value function is a convex
function. The new RAB method shows superior performance compared
to the other state-of-the-art general-rank RAB methods.
\end{abstract}

\section{Introduction}
It is  well known that when the desired signal is present in the
training data, the performance of adaptive beamforming methods
degrades dramatically in the presence of even a very slight
mismatch in the knowledge of the desired signal covariance matrix.
The mismatch between the presumed and actual source covariance
matrices occurs because of, for example, displacement of antenna
elements, time varying environment, imperfections of propagation
medium, etc. The main goal of any robust adaptive beamforming
(RAB) technique is to provide robustness against any such
mismatches.

Most of the RAB methods have been developed for the case of point
source signals when the rank of the desired signal covariance
matrix is equal to one \cite{Alex}-\cite{Arash2}. Among the
principles used for such RAB methods design are i) the worst-case
performance optimization \cite{Sergiy}-\cite{Li2}; ii)
probabilistic based performance optimization \cite{Sergiy2}; and
iii) estimation of the actual steering vector of the desired
signal \cite{Nasr}-\cite{Arash2}. In many practical applications
such as, for example, the incoherently scattered signal source or
source with fluctuating (randomly distorted) wavefronts, the rank
of the source covariance matrix is higher than one.  Although the
RAB methods of \cite{Alex}-\cite{Arash2} provide excellent
robustness against any mismatch of the underlying point source
assumption, they are not perfectly suited to the case when the
rank of the desired signal covariance matrix is higher than one.

The RAB for the general-rank signal model based on the explicit
modeling of the error mismatches has been developed in
\cite{Shahram} based on the worst-case performance optimization
principle. Although the RAB of \cite{Shahram} has a simple closed
form solution, it is overly conservative because the worst-case
correlation matrix of the desired signal may be negative-definite
\cite{Haihua1}-\cite{Haihua2}. Thus, less conservative approaches
have been developed in \cite{Haihua1}-\cite{Haihua2} by
considering an additional positive semi-definite (PSD) constraint
to the worst-case signal covariance matrix. The major shortcoming
of the RAB methods of \cite{Haihua1}-\cite{Haihua2} is that they
find only a suboptimal solution and there may be a significant gap
to the global optimal solution. For example, the RAB of
\cite{Haihua1} finds a suboptimal solution in an iterative way,
but there is no guarantee that such iterative method converges
\cite{Haihua2}. A closed-form approximate suboptimal solution is
proposed in \cite{LowComp}, however, this solution may be quite
far from the globally optimal one as well. All these shortcomings
motivate us to look for new efficient ways to solve the
aforementioned non-convex problem globally
optimally.\footnote{Some preliminary results have been presented
in \cite{ArashCamsap}.}

We propose a new method that is based on recasting the original
non-convex difference-of-convex functions (DC) programming problem
as the minimization of a one dimensional optimal value function.
Although the corresponding optimization problem of the newly
introduced optimal value function is non-convex, it can be
replaced with another equivalent function. The optimization
problem that corresponds to such new optimal value function is
convex and can be solved efficiently. The new one-dimensional
optimal value function is then minimized by the means of the newly
designed polynomial time DC (POTDC) algorithm (see also
\cite{POTDCconf}, \cite{POTDCjor}). We prove that the point found
by the POTDC algorithm for RAB for general-rank signal model with
positive semi-definite constraint is a Karush-Kuhn-Tucker (KKT)
optimal point. Moreover, we prove a number of results that lead us
to the equivalence between the claim of global optimality for the
POTDC algorithm as applied to the problem under consideration and
the convexity of the newly obtained one-dimensional optimal value
function. The latter convexity of the newly obtained
one-dimensional optimal value function can be checked numerically
by using the convexity on lines property of convex functions. The
fact that enables such numerical check is that the argument of
such optimal value function is proved to take values only in a
closed interval. In addition, we also develop tight lower-bound
for such optimal value function that is used in the simulations
for further confirming global optimality of the POTDC method.

The rest of the paper is organized as follows. System model and
preliminaries are given in Section II, while the problem is
formulated in Section III. The new proposed method is developed in
Section IV  followed by the simulation results in Section V.
Finally, Section VI presents our conclusions. This paper is
reproducible research and the software needed to generate the
simulation results will be provided to the IEEE Xplore together
with the paper upon its acceptance.

\section{System Model and Preliminaries}
The narrowband signal received by a linear antenna array with $M$
omni-directional antenna elements at the time instant $k$ can be
expressed as
\begin{equation} \label{signal_array}
\mathbf x (k)= \mathbf s (k) + \mathbf i (k) + \mathbf n (k)
\end{equation}
where $\mathbf s (k)$, $\mathbf i (k)$, and $\mathbf n (k)$ are
the statistically independent $M \times 1$ vectors of the desired
signal, interferences, and noise, respectively. The beamformer
output at the time instant $k$ is given as \vspace{-1mm}
\begin{equation} \label{output_signal}
y (k) = \mathbf w^H \mathbf x (k) \vspace{-1mm}
\end{equation}
where $\mathbf w$ is the $M \times 1$ complex beamforming vector
of the antenna array and $( \cdot )^H$ stands for the Hermitian
transpose. The beamforming problem is formulated as finding the
beamforming vector $\mathbf w$ which maximizes the beamformer
output signal-to-interference-plus-noise ratio (SINR) given as
\begin{equation} \label{SINR}
\mathrm {SINR}= \frac{\mathbf w^H {\mathbf R_{\rm s}} \mathbf
w}{\mathbf w^H {\mathbf R_{\rm i+n}} \mathbf w}
\end{equation}
where $\mathbf R_{\rm s} \triangleq E \{ \mathbf s (k) \mathbf s
(k)^H \} $ and $\mathbf R_{\rm i+n} \triangleq E \{ (\mathbf i (k)
+ \mathbf n (k)) (\mathbf i (k) + \mathbf n (k))^H \}$ are the
desired signal and interference-plus-noise covariance matrices,
respectively, and $E \{ \cdot \}$ stands for the statistical
expectation.

Depending on the nature of the desired signal source, its
corresponding covariance matrix can be of an arbitrary rank, i.e.,
$1 \leq {\rm rank}\{\mathbf R_{\rm s}\} \leq M$, where ${\rm
rank}\{\cdot\}$ denotes the rank operator. Indeed, in many
practical applications, for example, in the scenarios with
incoherently scattered signal sources or signals with randomly
fluctuating wavefronts, the rank of the desired signal covariance
matrix $\mathbf R_{\rm s}$ is greater than one \cite{Shahram}. The
only particular case in which, the rank of $\mathbf R_{\rm s}$ is
equal to one is the case of the point source.

The interference-plus-noise covariance matrix $\mathbf R_{\rm
i+n}$ is typically unavailable in practice and it is substituted
by the data sample covariance matrix
\vspace{-2mm}
\begin{equation} \label{datasample}
\hat{\mathbf R} = \frac{1}{K} \sum_{i=1}^K \mathbf x (i) \mathbf
x^H (i)
\end{equation}
where $K$ is number of the training data samples. The problem of
maximizing the SINR \eqref{SINR} (here we always use sample matrix
estimate $\hat{\mathbf R}$ instead of $\mathbf R_{\rm i+n}$) is
known as minimum variance distortionless response (MVDR)
beamforming and can be mathematically formulated as
\begin{eqnarray}
\min\limits_{{\bf w}}  {\bf w}^H \hat{\mathbf R} {\bf w} \quad
{\rm s.t.}   \quad {\bf w}^H  \mathbf R_{\rm s}  {\bf w} = 1.
\label{MVDR}
\end{eqnarray}
The solution to the MVDR beamforming problem \eqref{MVDR} can be
found as \cite{Alex}
\vspace{-2mm}
\begin{equation} \label{SMI} {\mathbf w}_{\rm SMI-MVDR}
= {\boldsymbol {\cal P}}\{\hat{\mathbf R}^{-1} \mathbf R_{\rm s}
\}
\end{equation}
which is known as the sample matrix inversion (SMI) MVDR
beamformer for general-rank signal model. Here $ {\boldsymbol
{\cal P}}\{\cdot\}$ stands for the principal eigenvector operator.

In practice, the actual desired signal covariance matrix $\mathbf
R_{\rm s}$ is usually unknown and only its presumed value is
available. The actual source correlation matrix can be modeled as
$ \mathbf R_{\rm s}= \tilde{\mathbf R}_{\rm s}+ \boldsymbol
\Delta_1$, where $\boldsymbol \Delta_1$ and $\tilde{\mathbf
R}_{\rm s}$ denote an unknown mismatch and the presumed
correlation matrices, respectively. It is well known that the MVDR
beamformer is very sensitive to such mismatches \cite{Shahram}.
RABs also address the situation when the sample estimate of the
data covariance matrix \eqref{datasample} is inaccurate (for
example, because of small sample size) and $\mathbf R =
\hat{\mathbf R}+\mathbf \Delta_2$, where $\mathbf  \Delta_2$ is an
unknown mismatch matrix to the data sample covariance matrix. In
order to provide robustness against  the norm-bounded mismatches
$\| \mathbf \Delta_1 \| \leq \epsilon $ and $\| \mathbf \Delta_2
\| \leq \gamma$ (here $\| \cdot \|$ denotes the Frobenius norm of
a matrix), the RAB of \cite{Shahram} uses the worst-case
performance optimization principle of \cite{Sergiy} and finds the
solution as
\begin{equation}
\mathbf w = {\boldsymbol {\cal P}} \{ (\hat{\mathbf R}+\gamma
\mathbf I)^{-1} (\tilde{\mathbf R}_{\rm s}-\epsilon \mathbf I)\}.
\label{closedform}
\end{equation}
Although the RAB of \cite{Shahram}  has a simple closed-form
solution \eqref{closedform}, it is overly conservative because the
constraint that the matrix $\tilde{\mathbf R}_{\rm s}+\boldsymbol
\Delta_1$ has to be positive semi-definite (PSD) is not considered
\cite{Haihua1}. For example, the worst-case desired signal
covariance matrix $\tilde{\mathbf R}_{\rm s}-\epsilon \mathbf I$
in \eqref{closedform} can be indefinite or negative definite if
$\tilde{\mathbf R}_{\rm s}$ is rank deficient.  Indeed, in the
case of incoherently scattered source, $\tilde{\mathbf R}_{\rm s}$
has the following form $ \tilde{\mathbf R}_{\rm s}=\sigma_{\rm
s}^2\int_{-\pi/2}^{\pi/2} \zeta (\theta) \mathbf a(\theta) \mathbf
a^H(\theta) d\theta $, where $ \zeta (\theta) $ denotes the
normalized angular power density, $\sigma_{\rm s}^2$ is the
desired signal power, and $\mathbf a(\theta)$ is the steering
vector towards direction $\theta$. For a uniform angular power
density on the angular bandwidth $\Phi$, the approximate numerical
rank of $\tilde{\mathbf R}_{\rm s}$ is equal to $(\Phi/\pi) \cdot
M$ \cite{Ali}. This leads to a rank deficient matrix
$\tilde{\mathbf R}_{\rm s}$ if the angular power density does not
cover all the directions. Therefore, the worst-case covariance
matrix $\tilde{\mathbf R}_{\rm s} - \epsilon \mathbf I$ is
indefinite or negative definite. Note that the worst-case data
sample covariance matrix $\hat{\mathbf R}+\gamma \mathbf I$ is
always positive definite.

\section{Problem Formulation}
Decomposing $\mathbf R_{\rm s}$ as $\mathbf R_{\rm s} = \mathbf
Q^H \mathbf Q$, the RAB problem for a norm bounded-mismatch $\|
\mathbf \Delta \| \leq \eta$ to the matrix $\mathbf Q$ is given as
\cite{Haihua1}
\begin{eqnarray}
&\min\limits_{\mathbf w}& \max_{ \| \mathbf \Delta_2 \| \leq
\gamma} \mathbf w^H(\hat{\mathbf R} + \mathbf \Delta_2)\mathbf w
\nonumber \\
&{\rm s.t.}& \min_{ \| \mathbf \Delta \| \leq \eta } \mathbf
w^H(\mathbf Q\!+\! \mathbf \Delta)^H (\mathbf Q\!+\! \mathbf
\Delta) \mathbf w \!\geq\! 1 . \label{PosconsProb2}
\end{eqnarray}
For every $\mathbf \Delta$ in the optimization problem
\eqref{PosconsProb2} whose norm is less than or equal to  $\eta$,
the expression $\mathbf w^H(\mathbf Q\!+\! \mathbf \Delta)^H
(\mathbf Q\!+\! \mathbf \Delta) \mathbf w \!\geq\! 1$  represents
a non-convex quadratic constraint with respect to $\mathbf w$.
Because there exists infinite number of mismatches $\mathbf
\Delta$, there also exists infinite number of such non-convex
quadratic constraints. By finding the minimum possible value of
the quadratic term $\mathbf w^H(\mathbf Q\!+\! \mathbf \Delta)^H
(\mathbf Q\!+\! \mathbf \Delta) \mathbf w$ with respect to
$\mathbf \Delta$ for a fixed $\mathbf w$, the infinite number of
such non-convex quadratic constraints can be replaced with a
single constraint. For this goal, we consider the following
optimization problem
\begin{eqnarray}
\min\limits_{\mathbf \Delta} && \mathbf w^H(\mathbf
Q\!+\! \mathbf \Delta)^H (\mathbf Q\!+\! \mathbf \Delta)
\mathbf w \nonumber \\
{\rm s.t.} &&\| \mathbf \Delta \|^2 \leq \eta^2 \label{subsub1}
\end{eqnarray}
where $\mathbf \Delta$ is a Hermitian matrix. This problem is
convex and its optimal value can be expressed as a function of
$\mathbf w$ as given by the following lemma.

\textbf{Lemma~1:} The optimal value of the optimization problem
\eqref{subsub1} as a function of $\mathbf w$ is equal to
\begin{eqnarray}
&& \min\limits_{ \| \mathbf \Delta \|^2 \leq \eta^2  }  \mathbf
w^H(\mathbf Q\!+\! \mathbf \Delta)^H (\mathbf Q\!+\! \mathbf
\Delta) \mathbf w  \nonumber
\\
&& =\left\{
\begin{array}{ll}
(\|\mathbf Q \mathbf w \| - \eta \|\ \! \mathbf w \|)^2, & \quad
\|\mathbf Q \mathbf w\| \geq  \eta  \| \mathbf w \|  \\
0 , & \quad\  \mbox{otherwise.}
\end{array}
\right. \label{Lemmaworstcasesol}
\end{eqnarray}

\begin{proof}
See Appendix, Subsection VII-A.
\end{proof}
The maximum of the quadratic term $\mathbf w^H(\hat{\mathbf R} +
\mathbf \Delta_2)\mathbf w$ with respect to $\mathbf \Delta_2$,
$\| \mathbf \Delta_2 \| \leq \gamma$ that appears in the the
objective of the problem \eqref{PosconsProb2} can be easily
derived as $ \mathbf w^H(\hat{\mathbf R} + \gamma \mathbf
I)\mathbf w$. It is obvious from \eqref{Lemmaworstcasesol} that
the desired signal can be totally removed from the beamformer
output if $\|\mathbf Q \mathbf w\| < \eta  \| \mathbf w \|$. Based
on the later fact, $\|\mathbf Q \mathbf w \| -  \eta \|\mathbf w
\|$ should be greater than or equal to zero. For any such $\mathbf
w$, the new constraint in the optimization problem \eqref{subsub1}
can be expressed as $\|\mathbf Q \mathbf w \| - \eta \|\ \mathbf w
\| \geq 1$. Since $\|\mathbf Q \mathbf w \| - \eta \|\ \! \mathbf
w \| \geq 1$ also implies that $\|\mathbf Q \mathbf w \| - \eta \|
\mathbf w \| \geq 0$, the RAB problem \eqref{PosconsProb2} can be
equivalently rewritten as
\begin{eqnarray}
\label{main_problem0} &\min\limits_{\mathbf w}& \, \mathbf w^H (
\hat{\mathbf {R}} + \gamma \mathbf I) \mathbf w \quad \nonumber \\
&{\rm s.t.}& \quad \|\mathbf Q \mathbf w \| - \eta \|\ \!\!
\mathbf w \|\geq 1 \; .
\end{eqnarray}
Due to the non-convex DC constraint, \eqref{main_problem0} is
non-convex DC programming problem \cite{POTDCconf},
\cite{POTDCjor}. {DC optimization problems are believed to be
NP-hard in general \cite{DC1}, \cite{DC2}. There is a number of
methods that can be applied to DC problems of type
\eqref{main_problem0} in the literature. Among these methods are
the generalized polyblock algorithm, the extended general power
iterative (GPI) algorithm \cite{We}, DC iteration-based method
\cite{DCiter}, etc. However, the existing methods do not guarantee
to find the solution of \eqref{main_problem0}, i.e., to converge
to the global optimum of \eqref{main_problem0} in polynomial time.
This means that the problem \eqref{main_problem0} is NP-hard. The
best what is possible to show, for example, for the DC
iteration-based method is that it can find a KKT optimal point.
The overall computational complexity of the DC iteration-based
method can be, however, quite high because the number of
iterations required to converge grows dramatically with the
dimension of the problem.}

Recently, the problem \eqref{main_problem0} has also been
suboptimally solved using an iterative semi-definite relaxation
(SDR)-based algorithm in \cite{Haihua1} which also does not result
in the globally optimal solution and for which the convergence
even to a KKT optimal point is not guaranteed. A closed-form
suboptimal solution for the aforementioned non-convex DC problem
has been also derived in \cite{LowComp}. Despite its computational
simplicity, the performance of the method of \cite{LowComp} may be
far from the global optimum and even the KKT optimal point.
Another iterative algorithm has been proposed in \cite{Haihua2},
but it modifies the problem \eqref{main_problem0} and solves the
modified problem instead which again gives no guarantees for
finding the globally optimal solution of the original problem
\eqref{main_problem0}. {In what follows, we develop a new
polynomial time algorithm for addressing the DC programming
problems of type \eqref{main_problem0}, prove that it finds at
least a KKT optimal point of the problem, and attempt to prove
that it actually solves the problem.}

\section{New Proposed Method}
In this section, we aim at solving the problem
\eqref{main_problem0} in a rigorous way, i.e., without using any
type of approximations. For this goal, we design a POTDC-type
algorithm (see also \cite{POTDCconf}, \cite{POTDCjor}) that can be
used for solving a class of DC programming problems in polynomial
time. More specifically, the POTDC algorithm can be efficiently
used for DC programming problems whose non-convex parts are
functions of only one variable. By introducing the auxiliary
optimization variable $\alpha \geq 1$ and setting $\|\mathbf Q
\mathbf w \| = \sqrt{\alpha}$, the problem \eqref{main_problem0}
can be equivalently rewritten as
%\begin{subequations}
%\label{main_problem_alpha}
\begin{eqnarray}
\min_{\mathbf w, \alpha}& &\, \mathbf w^H (\hat{\mathbf {R}} +
\gamma \mathbf  I) \mathbf w \quad \nonumber \\
{\rm s.t.}  && \mathbf w^H \mathbf Q^H \mathbf Q \mathbf w
= \alpha \label{firstconstraint} \nonumber\\
&&\mathbf w^H  \mathbf w \leq \frac{(\sqrt{\alpha}-1)^2}{\eta^2},
\quad \alpha \geq 1. \label{main_problem_alpha}
\end{eqnarray}
%\end{subequations}
Note that $\alpha$ is restricted to be greater than or equal to one
because $\|\mathbf Q \mathbf w \|$ is greater than or equal to one
due to the constraint of the problem \eqref{main_problem0}. For
future needs, we find the set of all $\alpha$'s for which the
optimization problem \eqref{main_problem_alpha} is feasible. Let
us define the following set for a fixed value of $\alpha \geq 1$,
\begin{equation}
S(\alpha)=\{\mathbf w\ | \ \mathbf w^H  \mathbf w \leq
(\sqrt{\alpha}-1)^2/\eta^2\}. \label{Set}
\end{equation}
It is trivial that for every $\mathbf w \in S(\alpha)$, the
quadratic term $\mathbf w^H \mathbf Q^H \mathbf Q \mathbf w$ is
non-negative as  $\mathbf Q^H \mathbf Q$ is a positive
semi-definite matrix. Using the minimax theorem \cite{Haykin}, it
can be easily verified that the maximum value of the quadratic
term $\mathbf w^H \mathbf Q^H \mathbf Q \mathbf w$ over $\mathbf w
\in S(\alpha)$ is equal to $\left((\sqrt{\alpha}-1)^2/\eta^2
\right) \cdot \lambda_{\rm max}\{\mathbf Q^H \mathbf Q\}$ and this
value is achieved by
\begin{equation}
\mathbf w_{\alpha}=\frac{\sqrt{\alpha}-1}{\eta} {\boldsymbol {\cal
P}} \{\mathbf Q^H \mathbf Q\}\ \in \ S(\alpha).
\end{equation}
Here $\lambda_{\rm max}\{\cdot\}$ stands for the largest
eigenvalue operator. Due to the fact that for any $0\leq \beta
\leq 1$, the scaled vector $\beta  \mathbf w_{\alpha} $ lies
inside the set $S(\alpha)$ \eqref{Set}, the quadratic term
$\mathbf w^H \mathbf Q^H \mathbf Q \mathbf w$ can take values only
in the interval $[0,\left( (\sqrt{\alpha}-1)^2/\eta^2 \right)
\cdot \lambda_{\rm max}\{\mathbf Q^H \mathbf Q\}]$ over $\mathbf w
\in S(\alpha)$.

Considering the later fact and also the optimization problem
\eqref{main_problem_alpha}, it can be concluded that $\alpha$ is
feasible if and only if $\alpha \in [0,\left(
(\sqrt{\alpha}-1)^2/\eta^2 \right) \cdot  \lambda_{\rm
max}\{\mathbf Q^H \mathbf Q\}] $ which implies that
\begin{equation}
\frac{(\sqrt{\alpha}-1)^2}{\eta^2} \cdot \lambda_{\rm
max}\{\mathbf Q^H \mathbf Q\} \geq \alpha \label{inequality}
\end{equation}
or, equivalently, that
\begin{equation}
\frac{(\sqrt{\alpha}-1)^2}{\alpha} \geq \frac{\eta^2}{\lambda_{\rm
max}\{\mathbf Q^H \mathbf Q\}}. \label{inequality2}
\end{equation}

The function $(\sqrt{\alpha}-1)^2 / \alpha$ is strictly increasing
and it is also less than or equal to one for $\alpha \geq 1$.
Therefore, it can be immediately found that the problem
\eqref{main_problem_alpha} is infeasible for any $\alpha \geq 1$
if $\lambda_{\rm max}\{\mathbf Q^H \mathbf Q\} \leq \eta ^2$.
Thus, hereafter, it is assumed that $\lambda_{\rm max}\{\mathbf
Q^H \mathbf Q\} > \eta ^2$. Moreover, using \eqref{inequality2}
and the fact that the function $(\sqrt{\alpha}-1)^2 / \alpha$ is
strictly increasing, it can be found that the feasible set of the
problem \eqref{main_problem_alpha} corresponds to
\begin{equation}
\alpha \geq \frac{1}{\left(1-\frac{\eta}{\sqrt{\lambda_{\rm
max}\{\mathbf Q^H \mathbf Q\}}}\right)^2} \geq 1.
\end{equation}

As we will see in the following sections, for developing the POTDC
algorithm for the problem \eqref{main_problem_alpha}, an
upper-bound for the optimal value of $\alpha$ in
\eqref{main_problem_alpha} is needed. Such upper-bound is obtained
in terms of the following lemma.

\textbf{Lemma~2:} The optimal value of the optimization variable
$\alpha$ in the problem \eqref{main_problem_alpha} is
upper-bounded by $\lambda_{\rm max}\left\{( \hat{\mathbf {R}} +
\gamma \mathbf I)^{-1} \mathbf Q^{H} \mathbf Q \right\} \cdot
\mathbf w_0^H (\hat{\mathbf {R}} + \gamma \mathbf I) \mathbf w_0$,
where $\mathbf w_0$ is any arbitrary feasible point of the problem
\eqref{main_problem_alpha}.
\begin{proof}
See Appendix, Subsection VII-B.
\end{proof}
Using Lemma~2, the problem \eqref{main_problem_alpha} can be
equivalently stated as
%\begin{subequations}
%\label{main_problem_alpha_new}
\begin{eqnarray}
&\min\limits_{ \theta_1 \leq \alpha \leq \theta_2 }&  \overbrace{
\min\limits_{\mathbf w}  \mathbf w^H (\hat{\mathbf {R}} + \gamma
\mathbf I) \mathbf w }^\text{Inner Problem} \nonumber \\
&{\rm s.t.}&  \mathbf w^H  \mathbf Q^H \mathbf Q \mathbf w \!=
\alpha ,\nonumber \\ & \ \ & \mathbf w^H  \mathbf w \!\leq\!
\frac{(\sqrt{\alpha}\!-\!1)^2}{\eta^2}
\label{main_problem_alpha_new}
\end{eqnarray}
%\end{subequations}
where
\begin{equation}
\theta_1= \frac{1}{{\left(1-\frac{\eta}{\sqrt{\lambda_{\rm
max}\{\mathbf Q^H \mathbf Q\}}}\ \right)^2}}
\end{equation}
and
\begin{equation}
\theta_2 =\lambda_{\rm max}\left\{( \hat{\mathbf {R}} + \gamma
\mathbf I)^{-1} \mathbf Q^{H} \mathbf Q \right\} \mathbf w_0^H
(\hat{\mathbf {R}} + \gamma \mathbf I) \mathbf w_0.
\end{equation}
It is easy to verify that $\theta_2 \geq \theta_1$. For a fixed
value of $\alpha$, the inner optimization problem in
\eqref{main_problem_alpha_new} is non-convex with respect to
$\mathbf w$. Based on the inner optimization problem in
\eqref{main_problem_alpha_new} when $\alpha$ is fixed, we define
the following {\it optimal value function}

\begin{eqnarray}
h(\alpha) \triangleq \Big \{ \min\limits_{\mathbf w} \mathbf w^H
(\hat{\mathbf {R}} \!+\! \gamma \mathbf I) \mathbf w \ | \ \mathbf
w^H  \mathbf Q^H \mathbf Q \mathbf w \!= \alpha,  \nonumber \\  \
\mathbf w^H  \mathbf w \!\leq\!
\frac{(\sqrt{\alpha}\!-\!1)^2}{\eta^2}  \Big \},  \quad \theta_1
\leq \alpha \leq \theta_2. \label{hdef1}
\end{eqnarray}

Using the optimal value function \eqref{hdef1}, the problem
\eqref{main_problem_alpha_new} can be equivalently expressed as
\begin{eqnarray}
\min\limits_{\alpha}&& h(\alpha) \quad {\rm s.t.}\quad \theta_1
\leq \alpha \leq \theta_2. \label{orgsimp0}
\end{eqnarray}

The corresponding optimization problem of $h(\alpha)$ for a fixed
value of $\alpha$ is non-convex. In what follows, we aim at
replacing $h(\alpha)$ with an equivalent optimal value function
whose corresponding optimization problem is  convex.

Introducing the matrix $\mathbf W \triangleq \mathbf w \mathbf
w^H$ and using the fact that for any arbitrary matrix $\mathbf A$,
$\mathbf w^H \mathbf {A} \mathbf w = {\rm tr} \{ \mathbf {A}
\mathbf w\mathbf w^H \}$  (here ${\rm tr}\{ \cdot \}$ stands for
the trace of a matrix), the function \eqref{hdef1} can be
equivalently recast as
\begin{eqnarray}
&&\!\!\!\!\!\!\!\!\!\!  h(\alpha) = \Big\{ \min\limits_{\mathbf W}
{\rm tr}\big\{ (\hat{\mathbf {R}} + \gamma \mathbf I) \mathbf W
\big\} \ |  \ {\rm tr}\{  \mathbf Q^H \mathbf Q \mathbf W\} =
\alpha, \nonumber \\ && {\rm tr}\{ \mathbf W\} \!\leq\!
\frac{(\sqrt{\alpha}\!-\!1)^2}{\eta^2}, \ \mathbf W \succeq
\mathbf 0,\     {\rm rank}\{\mathbf W\} =1  \Big \}, \nonumber \\
&&\quad \quad \theta_1 \leq \alpha \leq \theta_2
\end{eqnarray}
%where ${\rm rank}(\cdot)$ denotes the rank operator.
Dropping the rank-one constraint in the corresponding optimization
problem of $h(\alpha)$ for a fixed value of $\alpha$, $(\theta_1
\leq \alpha \leq \theta_2)$, a new optimal value function denoted
as $k(\alpha)$ can be defined as
\begin{eqnarray}
k(\alpha) &\triangleq& \Big \{ \min\limits_{\mathbf W} {\rm
tr}\big\{ (\hat{\mathbf {R}} + \gamma \mathbf I) \mathbf W \big\}
\ |  \ {\rm tr}\{  \mathbf Q^H \mathbf Q \mathbf W\} = \alpha,
\nonumber \\ &&  \ {\rm tr}\{ \mathbf W\} \!\leq\!
\frac{(\sqrt{\alpha}\!-\!1)^2}{\eta^2}, \ \mathbf W \succeq
\mathbf 0  \Big \}, \nonumber \\ && \theta_1 \leq \alpha \leq
\theta_2. \label{kalpha}
\end{eqnarray}

For brevity, we will refer to the optimization problems that
correspond to the optimal value functions $h(\alpha)$ and
$k(\alpha)$ when $\alpha$ is fixed, as the optimization problems
of $h(\alpha)$ and $k(\alpha)$, respectively. Note also that
compared to the optimization problem of $h(\alpha)$, the
optimization problem of $k(\alpha)$ is convex. More importantly,
the following lemma establishes the equivalence between the
optimal value functions $h(\alpha)$ and $k(\alpha)$.

\textbf{Lemma 3:} The optimal value functions $h(\alpha)$ and
$k(\alpha)$ are equivalent, i.e., $h(\alpha)=k(\alpha)$ for any
$\alpha \in [\theta_1, \theta_2]$. Furthermore, based on the
optimal solution of the optimization problem of $k(\alpha)$ when
$\alpha$ is fixed, the optimal solution of the optimization
problem of $h(\alpha)$ can be constructed.

\begin{proof}
See Appendix, Subsection VII-C.
\end{proof}

Based on Lemma~3, the original problem \eqref{orgsimp0} can be
expressed as
\begin{eqnarray}
\min\limits_{\alpha}&& k(\alpha) \quad {\rm s.t.} \quad \theta_1
\leq \alpha \leq \theta_2 \label{orgsimp}
\end{eqnarray}

It is noteworthy to mention that based on the optimal solution of
\eqref{orgsimp} denoted as $\alpha_{\rm opt}$, we can easily
obtain the optimal solution of the original problem
\eqref{orgsimp0} or, equivalently, the optimal solution of the
problem \eqref{main_problem_alpha_new}. Specifically, since the
optimal value functions $h(\alpha)$ and $k(\alpha)$ are
equivalent, $\alpha_{\rm opt}$ is also the optimal solution of the
problem \eqref {orgsimp0} and, thus, also the problem
\eqref{main_problem_alpha_new}. Moreover, the optimization problem
of $k(\alpha_{\rm opt})$ is convex and can be easily solved. In
addition, using the results in Lemma~3, based on the optimal
solution of the optimization problem of $k(\alpha_{\rm opt})$, the
optimal solution of the optimization problem of $h(\alpha_{\rm
opt})$ can be constructed. Therefore, in the rest of the paper, we
concentrate on the problem \eqref{orgsimp}.

Since for every fixed value of $\alpha$, the corresponding
optimization problem of $k(\alpha)$ is a convex semi-definite
programming (SDP) problem, one possible approach for solving
\eqref{orgsimp} is based on exhaustive search over $\alpha$. In
other words, $\alpha$ can be found by using an exhaustive search
over a fine grid on the interval of $[\theta_1,\theta_2]$.
Although this search method is inefficient, it can be used as a
benchmark.

Using the definition of the optimal value function $k(\alpha)$,
the problem \eqref{orgsimp} can be equivalently expressed as
%\begin{subequations}
%\label{main_problem_alpha_M}
\begin{eqnarray}
&\min\limits_{\mathbf W, \alpha}&\!\!\!\!\! {\rm tr} \left\{
\mathbf
{(\hat{\mathbf {R}} \!+\! \gamma \mathbf I)} \mathbf W \right \}
\quad \nonumber \\
&{\rm s.t.}& {\rm tr} \{ \mathbf Q^H \mathbf Q \mathbf W \} \!=\!
\alpha   \nonumber\\ && {\eta^2} {\rm tr} \{ \mathbf W \} \!\leq\!
(\sqrt{\alpha}
\!-\!1)^2 \nonumber \\
&& \mathbf W \succeq 0, \label{main_problem_alpha_M} \ \theta_1
\leq \alpha \leq \theta_2 .
\end{eqnarray}
%\end{subequations}
Note that replacing $h(\alpha)$ by $k(\alpha)$ results in a much
simpler problem. Indeed, compared to the original problem
\eqref{main_problem_alpha_new}, in which the first constraint is
non-convex, the corresponding first constraint of
\eqref{main_problem_alpha_M} is convex. All the constraints and
the objective function of the problem \eqref{main_problem_alpha_M}
are convex except for the constraint ${\rm tr} \{ \mathbf W \}
\leq (\sqrt{\alpha} - 1)^2 /\eta^2$ {which is non-convex only in a
single variable $\alpha$ and which makes the problem
\eqref{main_problem_alpha_M} non-convex overall. This single
non-convex constraint can be rewritten equivalently as $\eta^2
{\rm tr} \{ \mathbf W \} - (\alpha + 1) + 2\sqrt{\alpha} \leq 0$
where all the terms are linear with respect to $\mathbf W$ and
$\alpha$ except for the concave term of $\sqrt{\alpha}$. The
latter constraint can be handled iteratively by building a
POTDC-type algorithm (see also \cite{POTDCconf}, \cite{POTDCjor})
based on the iterative linear approximation of the non-convex term
$\sqrt{\alpha}$ around suitably selected points. It is interesting
to mention that this iterative linear approximation can be also
interpreted in terms of DC iteration over a single non-convex term
$\sqrt{\alpha}$. The fact that iterations are needed only over a
single variable helps to reduce dramatically the number of
iterations of the algorithm and allows for very simple
interpretations shown below.}

\subsection{Iterative POTDC Algorithm}
Let us consider the optimization problem
\eqref{main_problem_alpha_M} and replace the term $\sqrt{\alpha}$
by its linear approximation around $\alpha_c$, i.e.,
$\sqrt{\alpha} \approx \sqrt{\alpha_c} + (\alpha - \alpha_c)/(2
\sqrt{\alpha_c})$. It leads to the following SDP problem
%\begin{subequations}
%\label{relaxed}
\begin{eqnarray}
&\min\limits_{\mathbf W,\alpha}&\!\!\!\! {\rm tr} \left\{
(\hat{\mathbf {R}}
+ \gamma \mathbf I) \mathbf W \right\} \quad \nonumber \\
&{\rm s.t.}& \!\!\!\! {\rm tr} \{ \mathbf Q^H \mathbf Q \mathbf W
\} = \alpha  \nonumber \\
&&\!\!\!\!\eta^2 {\rm tr} \{ \mathbf W\! \}
\!+\!(\sqrt{\alpha_c}\!-\!1) \!+\! \alpha \left(\frac{1}{
\sqrt{\alpha_c}}\!-\!1\!\right) \leq \!0  \nonumber\\
&&\!\!\!\!  \mathbf W \succeq 0, \ \  \theta_1 \leq \alpha \leq
\label{relaxed} \theta_2.
\end{eqnarray}
%\end{subequations}
To understand  the POTDC algorithm intuitively and also to see how
the linearization points are selected in different iterations, let
us define the following optimal value function based on the
optimization problem \eqref{relaxed}
\begin{eqnarray}
\label{Ldef} l(\alpha,\alpha_c) \triangleq\!\!\!\!\!  &\Big
\{&\!\!\!\!\! \min\limits_{\mathbf W} {\rm tr}\big\{ (\hat{\mathbf
{R}} + \gamma \mathbf I)  \mathbf W \big\} \ |  \ {\rm tr}\{
\mathbf Q^H \mathbf Q \mathbf W\} = \alpha, \nonumber \\ && \eta^2
{\rm tr} \{ \mathbf W\! \} \!+\!(\sqrt{\alpha_c}\!-\!1) \!+\!
\alpha \left(\frac{1}{\sqrt{\alpha_c}}\!-\!1\!\right) \leq \!0,
\nonumber\\ && \mathbf W \succeq \mathbf 0  \Big \},  \ \ \ \
\theta_1 \leq \alpha \leq \theta_2  .
\end{eqnarray}
where $\alpha_c$ in $l(\alpha,\alpha_c)$ denotes the linearization
point. The optimal value function $l(\alpha,\alpha_c)$ can be also
obtained through $k(\alpha)$ in \eqref{kalpha} by replacing the
term $\sqrt{\alpha}$ in $\eta^2 {\rm tr} \{ \mathbf W \} - (\alpha
+ 1) + 2\sqrt{\alpha} \leq 0$ with its linear approximation around
$\alpha_c$. Since  $\sqrt{\alpha}$ and its linear approximation
have the same values at $\alpha_c$, $l(\alpha,\alpha_c)$ and
$k(\alpha)$ take the same values at this point. The following
lemma establishes the relationship between the optimal value
functions $k(\alpha)$ and $l(\alpha,\alpha_c)$.

{\bf Lemma~4: } $l(\alpha,\alpha_c)$ is a convex upper-bound of
$k(\alpha)$ for any arbitrary $\alpha_c \in [\theta_1,\theta_2]$,
i.e., $l(\alpha,\alpha_c) \geq k(\alpha)$, $\forall \alpha \in
[\theta_1,\theta_2]$ and  $l(\alpha,\alpha_c)$ is convex with
respect to $\alpha$. Furthermore, the values of the optimal value
functions $k(\alpha)$ and $l(\alpha,\alpha_c)$ as well as their
right and left derivatives are equal at this point. In other
words, under the condition that $k(\alpha)$ is differentiable at
$\alpha_c$, $l(\alpha,\alpha_c)$ is tangent to $k(\alpha)$ at this
point.

\begin{proof}
See Appendix, Subsection VII-D.
\end{proof}
In what follows, for the sake of clarity of the explanations, it
is assumed that the function  $k(\alpha)$ is differentiable over
the interval of $(\theta_1,\theta_2)$, however, this property is
not generally required as we will see later. Let us consider an
arbitrary point, denoted as $\alpha_0$, $\alpha_0 \in
(\theta_1,\theta_2)$ as the initial linearization point, i.e.,
$\alpha_c=\alpha_0$. Based on Lemma~4, $l(\alpha,\alpha_0)$ is a
convex function with respect to $\alpha$ which is tangent to
$k(\alpha)$ at the linearization point $\alpha=\alpha_0$, and it
is also an upper-bound to $k(\alpha)$. Let $\alpha_1$ denote the
global minimizer of $l(\alpha,\alpha_0)$ that can be easily
obtained due to the convexity of $l(\alpha,\alpha_0)$ with
polynomial time complexity.

\begin{figure}[h]
\begin{center}
 \includegraphics[scale=.3]{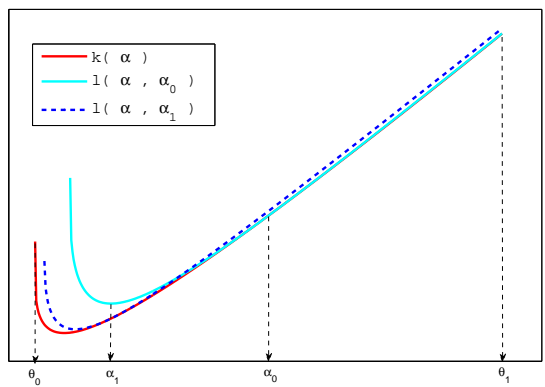} \hspace{-15mm}
\caption{Iterative method for minimizing the optimal value
function $k(\alpha)$. The convex optimal value function
$l(\alpha,\alpha_0)$ is an upper bound to $k(\alpha)$ which is
tangent to it at $\alpha=\alpha_0$, and its minimum is denoted as
$\alpha_1$. The point $\alpha_1$ is used to establish another
convex upper-bound function denoted as  $l(\alpha,\alpha_1)$ and
this proces continutes.} \label{IterFig}
\end{center}
\end{figure}

Since $l(\alpha,\alpha_0)$ is tangent to $k(\alpha)$ at
$\alpha=\alpha_0$ and it is also an upper-bound for $k(\alpha)$,
it can be concluded that $\alpha_1$ is a decent point for
$k(\alpha)$, i.e., $k(\alpha_1) \leq k(\alpha_0)$ as it is shown
in Fig.\ref{IterFig}. Specifically, the fact that
$l(\alpha,\alpha_0)$ is tangent to $k(\alpha)$ at
$\alpha=\alpha_0$ and $\alpha_1$ is the global minimizer of
$l(\alpha,\alpha_0)$ implies that
\begin{equation}
l(\alpha_1,\alpha_0) \leq l(\alpha_0,\alpha_0)=k(\alpha_0).
\label{eq1}
\end{equation}
Furthermore, since $l(\alpha,\alpha_0)$ is an upper-bound for
$k(\alpha)$, it can be found that $k(\alpha_1) \leq
l(\alpha_1,\alpha_0)$. Due to the later fact and also the equation
\eqref{eq1}, it is concluded that $k(\alpha_1) \leq k(\alpha_0)$.

Choosing $\alpha_1$ as the linearzation point in the second
iteration, and finding the global minimizer of
$l(\alpha,\alpha_1)$ over the interval $[\theta_1,\theta_2]$
denoted as $\alpha_2$, another decent point can be obtained, i.e.,
$k(\alpha_2) \leq k(\alpha_1)$. This process can be continued
until convergence.

Then the proposed iterative decent method can be described as
shown in Algorithm~1. The following lemma about the convergence of
Algorithm~1 and the optimality of the solution obtained by this
algorithm is in order. Note that this lemma makes no assumptions
about the differentiability of the optimal value function
$k(\alpha)$.

\textbf{Lemma~5:} The following statements regarding Algorithm~1
are true:
\begin{description}
\item[i)] The optimal value of the optimization problem in
Algorithm~1 is non-increasing over iterations, i.e.,
\begin{eqnarray}
&&\!\!\!\!\!\!\!\!\!\!\!\!{\rm tr}\! \left\{ ( \hat{\mathbf {R}} +
\gamma \mathbf I) \mathbf W_{opt,i+1} \right\} \leq {\rm tr}\!
\left\{ ( \hat{\mathbf {R}} + \gamma \mathbf I) \mathbf W_{opt,i}
\right\}, \nonumber  \\ && \ i \geq 1. \nonumber
\end{eqnarray}

\item[ii)]  Algorithm~1 converges.
\item[iii)] Algorithm~1 converges to a KKT solution, i.e., a
solution which satisfies the KKT optimality conditions.
\end{description}

\begin{proof}
See Appendix, Subsection VII-E.
\end{proof}

\begin{algorithm}[h]                   % enter the algorithm environment
\caption{The iterative POTDC algorithm}% give the algorithm a caption
\label{alg1}                           % and a label for \ref{} commands later in the document
\begin{algorithmic}                    % enter the algorithmic environment
\REQUIRE An arbitrary $\alpha_{c}  \in [\theta_1, \theta_2]$, \\
\quad \quad \ \ \ the termination threshold $\zeta$, \\
\quad \quad \ \ \ set $i$ equal to 1. \REPEAT
  \STATE    Solve the following optimization problem using $\alpha_c$
  to obtain $\mathbf W_{opt}$ and $\alpha_{opt}$
  \begin{eqnarray}
&\min\limits_{\mathbf W,\alpha}&\!\!\!\! {\rm tr} \left\{
(\hat{\mathbf {R}}
+ \gamma \mathbf I) \mathbf W \right\} \quad \nonumber \\
&{\rm s.t.}& \!\!\!\! {\rm tr} \{ \mathbf Q^H \mathbf Q
\mathbf W \} = \alpha  \nonumber \\
&&\!\!\!\!\eta^2 {\rm tr} \{ \mathbf W\! \}
\!+\!(\sqrt{\alpha_c}\!-\!1) \!+\!
\alpha \left(\frac{1}{\sqrt{\alpha_c}}\!-\!1\!\right) \leq \!0
\nonumber \\
&&\!\!\!\! \mathbf W \succeq 0, \ \  \theta_1 \leq \alpha \leq
\theta_2 \nonumber
\end{eqnarray}
and set \STATE \quad \quad \ \ \ $\mathbf W_{opt,i} \leftarrow
\mathbf W_{opt}$, \ \    $\alpha_{opt,i}\leftarrow   \alpha_{opt}$
\STATE  \quad \quad \ \ \ \ \ \ \ \ $\alpha_{c}\leftarrow
\alpha_{opt}$, \ \ \  $i \leftarrow  i+1 $
  \UNTIL{$\ $} \\
  \STATE  \ \ \ \  ${\rm tr} \left\{ (\hat{\mathbf {R}}
+ \gamma \mathbf I) \mathbf W_{opt,i-1} \right\} \!-\! {\rm tr}
\left\{ (\hat{\mathbf {R}} + \gamma \mathbf I) \mathbf W_{opt,i}
\right\} \leq \zeta$ for $i \geq 2$ .
\end{algorithmic}
\end{algorithm}

{There is a great evidence that the point obtained by Algorithm~1
(POTDC algorithm) is also the globally optimal point. This
evidence is based on the following observation. The optimal value
function $k(\alpha)$ of \eqref{kalpha} is a convex function with
respect to $\alpha$. This observation is supported by numerous
checks of convexity of $k(\alpha)$ for any arbitrary positive
semi-definite matrices $\hat{\mathbf R}$ and $\mathbf R_{\rm
s}=\mathbf Q^H \mathbf Q$. Such numerical checks are performed
using the convexity on lines property of convex functions, and
they are possible because $\alpha$ takes values only from a closed
interval as it is shown before. As a result, if the optimal value
function $k(\alpha)$ of \eqref{kalpha} is a convex function of
$\alpha$, the proposed POTDC algorithm achieves the global optimal
solution (see Fig.~1 and corresponding explanations to how the
POTDC algorithm works). The following formal conjecture is then in
order.}

\textbf{Conjecture~1:} For any arbitrary positive semi-definite
matrices $\hat{\mathbf R}$ and $\mathbf R_{\rm s}=\mathbf Q^H
\mathbf Q$ and positive values of $\gamma$ and $\eta$, the optimal
value function $k(\alpha)$ defined in \eqref{kalpha} is a convex
function of $\alpha \in [\theta_1,\theta_2]$.

{It is worth noting that even a more relaxed property of the
optimal value function $k(\alpha)$ would be sufficient to
guarantee global optimality for the POTDC algorithm. Specifically,
if $k(\alpha)$ defined in \eqref{kalpha} is a strictly
quasi-convex function of $\alpha \in [\theta_1,\theta_2]$, then
the point found by the POTDC algorithm will be the global optimum
of the optimization problem \eqref{main_problem0}. The evidence,
however, is even more optimistic as stated in Conjecture~1 that
\eqref{kalpha} is a convex function of $\alpha \in
[\theta_1,\theta_2]$.} The computational complexity of Algorithm~1
is equal to that of the SDP optimization problem in Algorithm~1,
that is, ${\cal O}((M+1)^{3.5})$ times the number of iterations
(see also Simulation example~1 in the next section). The RAB
algorithm of \cite{Haihua1} is iterative as well and its
computational complexity is equal to ${\cal O}(M^{3.5})$ times the
number of iterations. The complexity of the RABs of \cite{Shahram}
and \cite{LowComp} is ${\cal O}(M^{3})$. {The comparison of the
overall complexity of the proposed POTDC algorithm with that of
the DC iteration-based method will be explicitly performed in
Simulation example~3.} Although the computational complexity of
the new proposed method may be slightly higher than that of the
other RABs, it finds the global optimum and results in superior
performance as it is shown in the next section.

\subsection {Lower-Bounds for the Optimal Value}
We also aim at developing a tight lower-bound for the optimal
value of the optimization problem \eqref{main_problem_alpha_M}.
Such lower-bound can be used for assessing the performance of the
proposed iterative algorithm.

As it was mentioned earlier, although the objective function of
the optimization problem \eqref{main_problem_alpha_M} is convex,
its feasible set is non-convex due to the second constraint of
\eqref{main_problem_alpha_M}. A lower-bound for the optimal value
of \eqref{main_problem_alpha_M} can be achieved by replacing the
second constraint of  \eqref{main_problem_alpha_M} by its
corresponding convex-hull. However, such lower-bound may not be
tight. In order to obtain a tight lower-bound, we can divide the
sector $[\theta_1,\theta_2]$ into $N$ subsector and solve the
optimization problem \eqref{main_problem_alpha_M} over each
subsector in which the second constraint of
\eqref{main_problem_alpha_M} has been replaced with the
corresponding convex hull. The minimum of the optimal values of
such optimization problem over the subsectors is the lower-bound
for the problem \eqref{main_problem_alpha_M}. It is obvious that
by increasing $N$, the lower-bound becomes tighter.

\section{Simulation Results}
Let us consider a uniform linear array (ULA) of $10$
omni-directional antenna elements with the inter-element spacing
of half wavelength. Additive noise in antenna elements is modeled
as spatially and temporally independent complex Gaussian noise
with zero mean and unit variance. Throughout all simulation
examples, it is assumed that in addition to the desired source, an
interference source with the interference-to-noise ratio (INR) of
$10$~dB impinges on the antenna array. For obtaining each point in
the simulation examples, $100$ independent runs are used {unless
otherwise is specified} and the sample data covariance matrix is
estimated using $K= 20$ snapshots.

The new proposed method is compared, in terms of the output SINR
to the RAB methods of  \cite{Shahram}, \cite{Haihua1}, and
\cite{LowComp}. The proposed method, the method of \cite{Haihua1}
(the best among previous methods), and the lower-bound on the
objective value of the problem \eqref{main_problem0} are also
compared in terms of the achieved values for the objective.

The diagonal loading parameters of $\gamma = 10$ and $\eta = 0.3
\sqrt{{\rm tr} \{ \mathbf R_{\rm s} \}}$ are chosen for all the
aforementioned methods. The initial $\alpha_0$ in the first
iteration of the proposed POTDC method equals to $(\theta_1 +
\theta_2)/2$ {unless otherwise is specified}. The termination
threshold $\zeta$ for the proposed algorithm is chosen to be equal
to $10^{-6}$.

\subsection{Simulation Example~1}
In this example, the desired and interference sources are locally
incoherently scattered with Gaussian and uniform angular power
densities with central angles of $30^\circ$ and $10^\circ$,
respectively. Both sources have the same angular spread of
$4^\circ$. The presumed knowledge of the desired source is
different from the actual one and is characterized by an
incoherently scattered source with Gaussian angular power density
whose central angle and angular spread are $32^\circ$ and
$1^\circ$, respectively. Note that, the presumed knowledge about
the shape of the angular power density of the desired source is
correct while the presumed central angle and angular spread
deviate from the actual one.

\begin{figure}[t]
\begin{center}
\includegraphics[scale=.4]{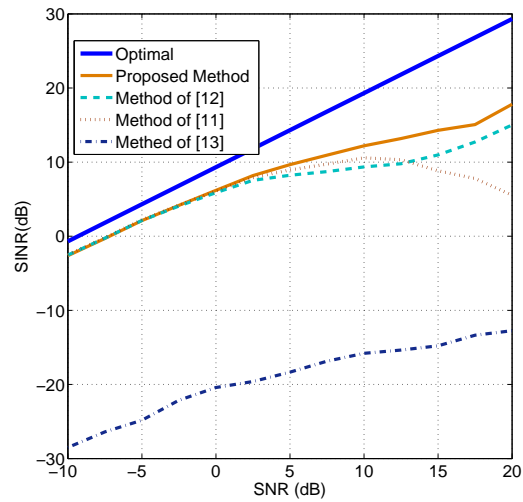}
\caption{Example 1: Output SINR versus SNR.} \label{Fig1}
\end{center}
\end{figure}

\begin{figure}[t]
\begin{center}
\includegraphics[scale=.4]{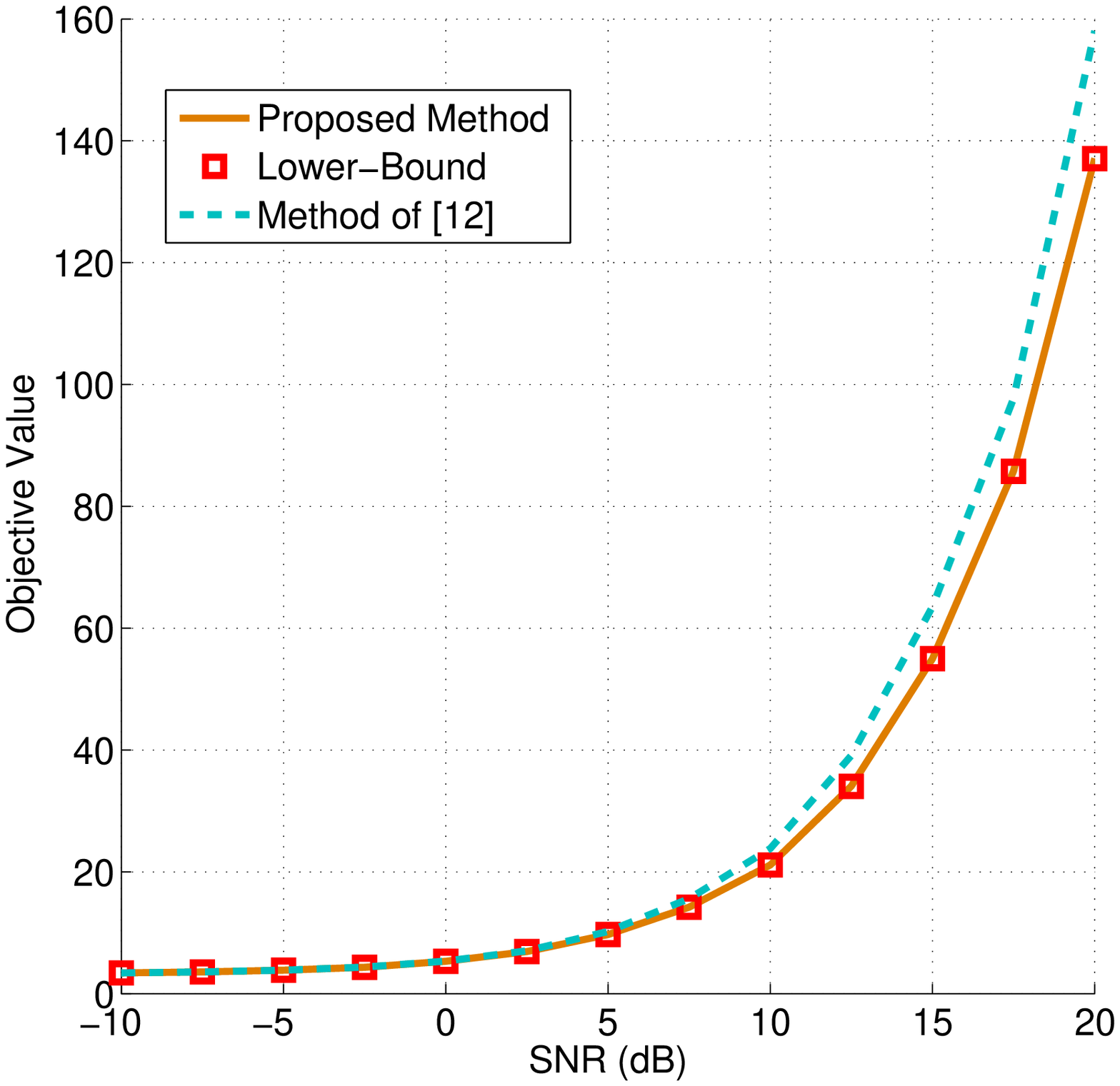}
\caption{Example 1: Objective function of the problem
\eqref{main_problem0} versus SNR.} \label{Fig2}
\end{center}
\end{figure}

\begin{figure}[t]
\begin{center}
\includegraphics[scale=.4]{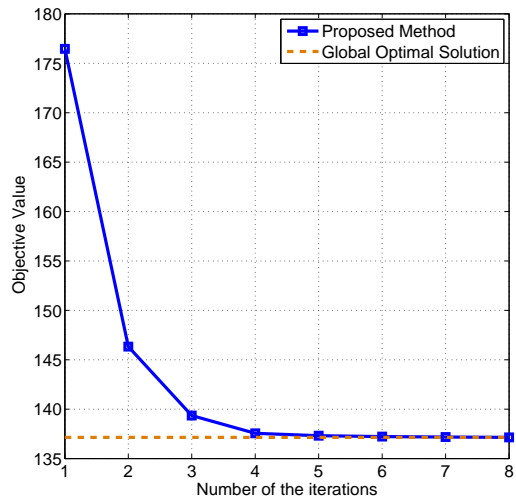}
\caption{Example 1: Objective function of the problem
\eqref{main_problem0} versus the number of iterations.}
\label{Convergence}
\end{center}
\end{figure}

In Figs.~\ref{Fig1} and \ref{Fig2}, the output SINR and the
objective function values of the problem \eqref{main_problem0},
respectively, are plotted versus SNR. It can be observed from the
figures that the proposed new method based on the POTDC algorithm
has superior performance over the other RABs. Although the method
of \cite{Haihua1} does not have a guaranteed convergence, it
results in a better average performance as compared to the method
of \cite{Shahram} and \cite{LowComp}. Moreover, the
Fig.~\ref{Fig2} confirms that the new proposed method archives the
global minimum of the optimization problem \eqref{main_problem0}
since the corresponding objective value coincides with the
lower-bound on the objective function of the problem
\eqref{main_problem0}. Fig.~\ref{Convergence} shows the
convergence of the iterative POTDC method. It shows the average of
the optimal value found by the algorithm over iterations for
$SNR=20$ dB. It can be observed that the proposed algorithm
converges to the global optimum in about 4 iterations.

\subsection{Simulation Example~2}
In this example, we also consider the locally incoherently
scattered desired and interference sources. However, compared to
the previous example, there is a substantial error in the
knowledge of the desired source angular power density.

\begin{figure}[t]
\begin{center}
\includegraphics[scale=.4]{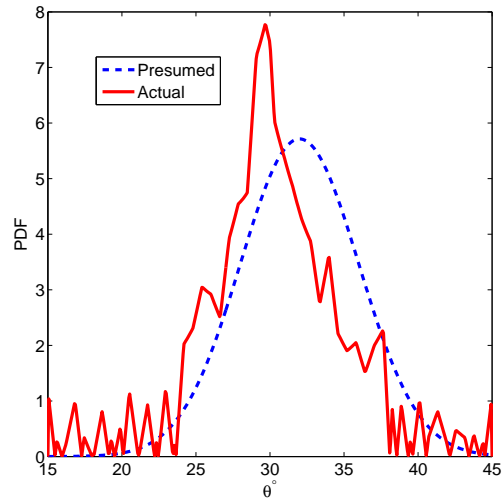}
\caption{Example 2: Actual and presumed angular power densities of
general-rank source.} \label{Fig3}
\end{center}
\end{figure}

The interference source is modeled as in the previous example,
while the angular power density of the desired source is assumed
to be a truncated Laplacian function distorted by severe
fluctuations. The central angle and the scale parameter of the
Laplacian distribution is assumed to be $30^\circ$ and $0.1$,
respectively, and it is assumed to be equal to zero outside of the
interval $[15^\circ,45^\circ]$ as it has been shown in Fig.
\ref{Fig3}. The presumed knowledge of the desired source is
different from the actual one and is characterized by an
incoherently scattered source with Gaussian angular power density
whose central angle and angular spread are $32^\circ$ and
$1^\circ$, respectively.

\begin{figure}[t]
\begin{center}
\includegraphics[scale=.4]{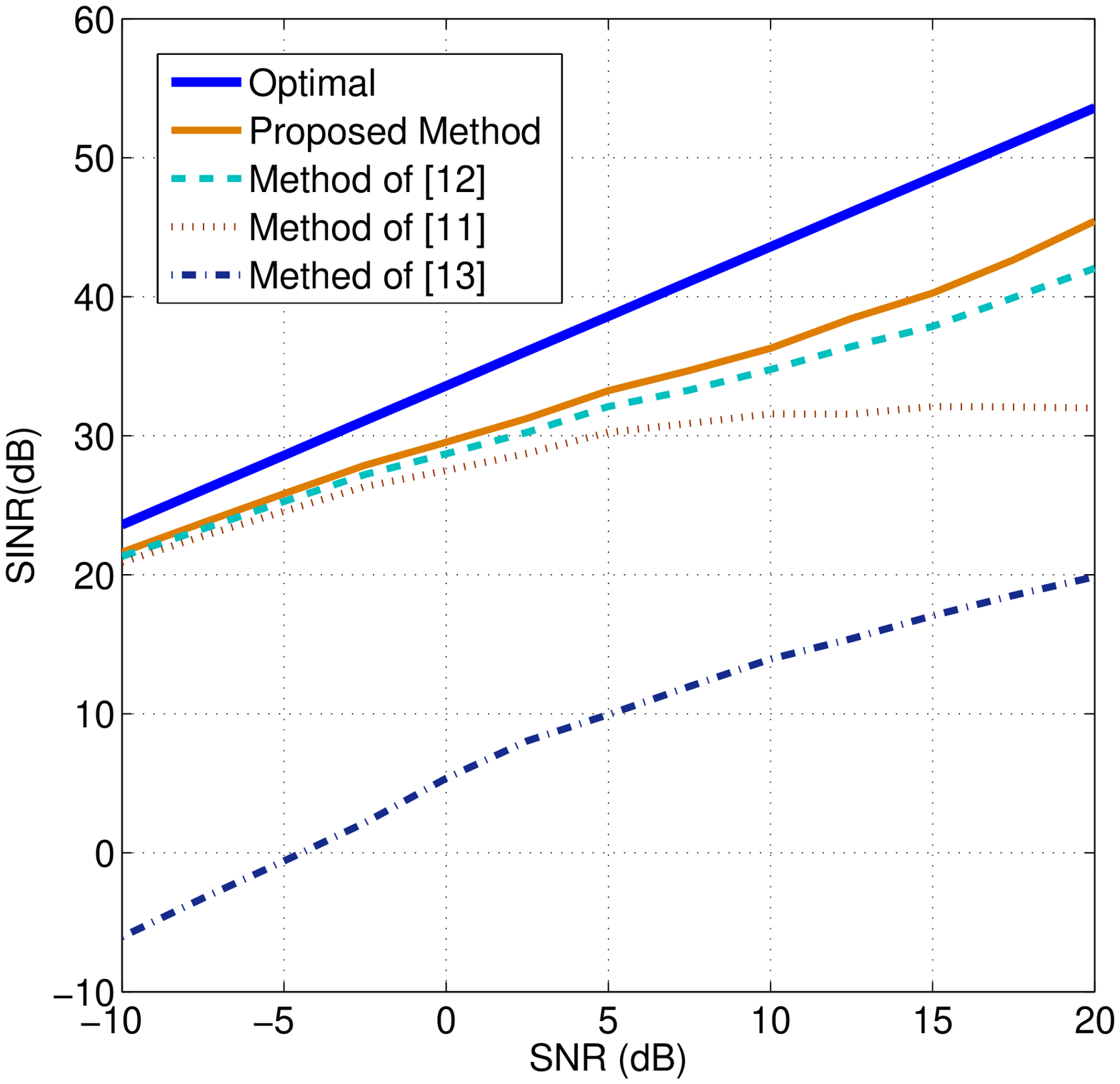}
\caption{Example 2: Output SINR versus SNR.} \label{Fig4}
\end{center}
\end{figure}

\begin{figure}[t]
\begin{center}
\includegraphics[scale=.4]{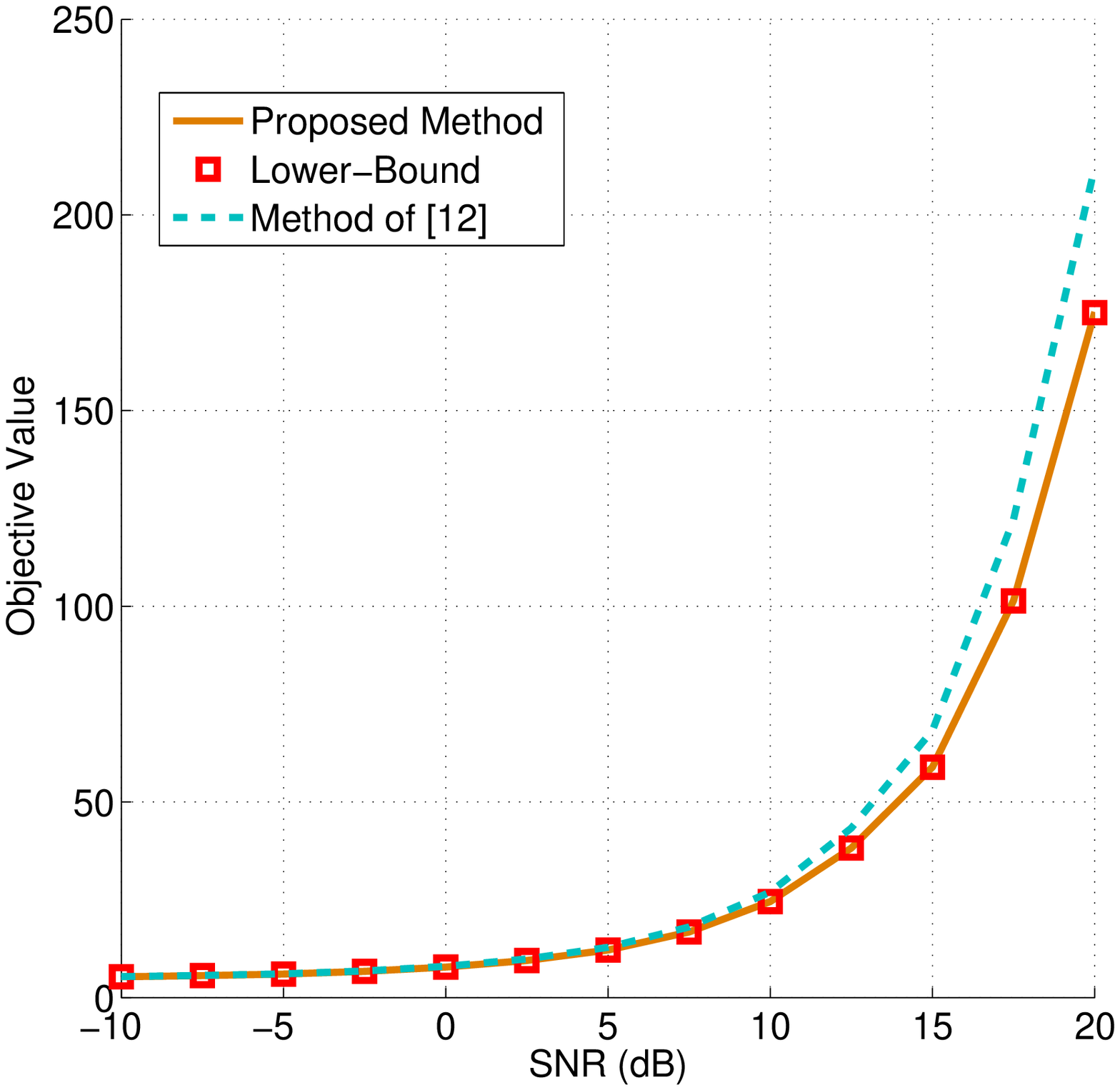}
\caption{Example 2: Objective function of the problem
\eqref{main_problem0} versus SNR.} \label{Fig5}
\end{center}
\end{figure}

Figs.~\ref{Fig4} and \ref{Fig5} depict the corresponding output
SINR and the objective function values of the problem
\eqref{main_problem0} obtained by the beamforming methods tested
versus SNR. Form these figures, it can be concluded that the
proposed new method has superior performance over the other
methods as well as it achieves the global minimum of the
optimization problem \eqref{main_problem0}

{
\subsection{Simulation Example~3}
In this example, we compare the efficiency of the proposed POTDC
method to that of the DC iteration-based method that can be
written for the problem under consideration as
\begin{eqnarray}
\!\!\!\!\!\!\!&  \min\limits_{\mathbf w}& \mathbf w^H(\mathbf
{\hat R} + \gamma \mathbf I) \mathbf w  \nonumber \\
\!\!\!\!\!\!\!&{\rm s.t.}& \!\!f(\mathbf w^{(k)})\!+\! \langle
\bigtriangledown f( \mathbf w^{(k)}, \mathbf w-\mathbf
w^{(k)})\rangle \!-\! \eta \| \mathbf w \|\! \geq\! 1
\label{DC-iteration}
\end{eqnarray}
where the function $f(\mathbf w) = \| \mathbf Q \mathbf w \| $ is
replaced with the first two terms of the Taylor expansion of
$f(\mathbf w)$ around $\mathbf w^{(k)}$. At the first iteration
$\mathbf w^{(1)}$ is initialized and in the next iterations
$\mathbf w^{(k)}$ is selected as the optimal $\mathbf w$ obtained
from solving \eqref{DC-iteration} in the previous iteration. Thus,
the iteration are performed over the whole vector of variables of
the problem.}

{The simulation set up is the same as in our Simulation example~1
except that different number of antennas are used. For a fair
comparison, the initial point $\alpha_0$ in the proposed POTDC
method and $\mathbf w^{(1)}$ in \eqref{DC-iteration} are chosen
randomly. Table~I shows the average number of the iterations for
the aforementioned methods versus the size of the antenna array.
The accuracy is set to $10^{-6}$, $SNR=-10$~dB, and each number in
the table is obtained by averaging the results from 200 runs. From
this table, it can be seen that the number of the iterations for
the POTDC method is almost fixed while it increases for the
DC-iteration method as the size of the array increases. The latter
phenomenon can be justified by considering the DC iteration-type
interpretation of the POTDC method over the one dimensional
function of $k(\alpha)$. The dimension of $k(\alpha)$ is
independent of the size of the array (thus, the size of the
optimization problem), while the size of search space for the DC
iteration-based method \eqref{DC-iteration}, that is, $2M$,
increases as the size of the array increases.
\begin{table}[h]
\begin{center}
\caption{Average number of the iterations}
\begin{tabular}{ | p{1.4cm} | p{0.55cm} | p{0.55cm} | p{0.55cm} |p{0.55cm}|p{0.65cm}|p{0.65cm}|p{0.65cm}|}
\hline
Array size & 8  & 10 & 12 & 14 & 16 & 18 & 20\\
\hline POTDC &2.940& 2.855&  2.805& 2.835&  2.870  & 2.840&  2.920  \\
\hline DC iteration  &5.930&6.925&7.870&9.180&10.430&11.890&13.305 \\ \hline
     \end{tabular}
\end{center}
\end{table}
The average (over 200 runs) CPU time for the aforementioned
methods is also compared in Table~2. Both methods have been
implemented in Matlab using CVX software and run on the same
desktop with Pentium(R) 4 CPU 3.40~GHz.
\begin{table}[h]
\begin{center}
\caption{Average CPU time}
\begin{tabular}{ | p{1.4cm} | p{0.55cm} | p{0.55cm} | p{0.55cm} |p{0.55cm}|p{0.55cm}|p{0.55cm}|p{0.65cm}|}
\hline
Array size & 8  & 10 & 12 & 14 & 16 & 18 & 20\\
\hline POTDC & 0.851 & 0.867 & 0.939 & 1.056 & 1.153 & 1.269 & 1.403    \\
\hline DC iteration  & 4.353& 3.897& 5.882&5.366& 7.870& 8.575& 10.041\\ \hline
     \end{tabular}
\end{center}
\end{table}
Table~2 confirms that the proposed method is more efficient that
the DC iteration-based one in terms of the time which is spent for
solving the same problem. Note that although the number of
variables in the matrix $\mathbf W$ of the optimization problem
\eqref{relaxed} is in general $M^2 + 1$ (since $\mathbf W$ has to
be a Hermitian matrix) after the rank one constraint is relaxed,
the probability that the optimal $\mathbf W$ is rank one has been
shown to be very high \cite{Arash2}, \cite{TomMag}-\cite{Tom2007}.
It is also approved by this our simulations. Thus, in almost all
cases, for different data sets, the actual dimension of the
problem \eqref{relaxed} is $2M + 1$. As a result, the average
complexity of solving \eqref{relaxed} is significantly smaller
than the worst-case complexity.}

\section{Conclusion}
We have considered the RAB problem for general-rank signal model
with additional positive semi-definite constraint. Such RAB
problem corresponds to a non-convex DC optimization problem. We
have studied this non-convex DC problem rigorously and designed
the POTDC algorithm for solving it. {It has been proved that the
point found by the POTDC algorithm for RAB for general-rank signal
model with positive semi-definite constraint is a KKT optimal
point. Moreover, there is a strong evidence that the POTDC method
actually finds the globally optimal point of the problem under
consideration, which is shown in terms of a number of lemmas and
one conjecture. Specifically, we have proved a number of results
that lead us to the equivalence between the claim of global
optimality for the POTDC algorithm as applied to the problem under
consideration and the convexity of the one-dimensional optimal
value function \eqref{kalpha}. The latter convexity has been
checked numerically by using the convexity on lines property of
convex functions. The fact that enables such numerical check is
that the argument of this one-dimensional optimal value function
is proved to take values only in a closed interval. The resulted
RAB method shows superior performance compared to the other
existing methods in terms of the output SINR. It also has lower
average overall complexity than the other traditional methods that
can be used for the same optimization problem such as, for
example, the DC iteration-based method. None of the existing
methods used for DC programming problems, however, guarantee that
the problem can be solved in polynomial time. Thus, the
fundamental development of this work is that the claim of global
optimality of the proposed POTDC method boils down to one
conjecture that can be easily checked numerically. It implies that
certain relatively simple DC programming problems (an example of
such problem is the RAB for general-rank signal model with
additional positive semi-definite constraint), which are believed
to be NP-hard, are likely not NP-hard.}

\section{Appendix}
\subsection{Proof of Lemma 1}
Since the objective function as well as the  constraints of the
optimization problem \eqref{subsub1} are all quadratic functions
of $\mathbf \Delta$, this problem is convex. It is easy to verify
that this problem satisfies the Slater's constraint qualification
and as a result the KKT conditions are necessary and sufficient
optimality conditions. Let us introduce the Lagrangian as
\begin{eqnarray}
L(\mathbf \Delta,\mu)\!\!\!\!&=&\!\!\!\!\mathbf w^H \left(\mathbf
Q^H \mathbf Q + \mathbf Q^H \mathbf \Delta + \mathbf \Delta^H
\mathbf Q + \mathbf \Delta^H \mathbf \Delta \right)\mathbf w
\nonumber \\ && +\ \mu ( \| \mathbf \Delta \|^2 - \eta^2 )
\end{eqnarray}
where $\mu$ is the non-negative Lagrange multiplier. The KKT
optimality conditions are
\begin{subequations}
\begin{eqnarray}
&&\boldsymbol \nabla_{\mathbf \Delta} L(\mathbf \Delta,\mu)
= \mathbf 0 \label{zerograd} \\
&&\| \mathbf \Delta \|^2 \leq \eta^2  \label{KKT1}\\
&&\mu ( \| \mathbf \Delta \|^2 - \eta^2 ) =0  \label{KKT2} \\
&&\mu \geq 0. \label{KKT3}
\end{eqnarray}
\end{subequations}
where $\mathbf 0$ is the vector of zeros. Using the matrix
differentiation, the zero gradient condition \eqref{zerograd} can
be expressed as $\mathbf Q \mathbf w \mathbf w^H + \boldsymbol
\Delta \mathbf w \mathbf w^H + \mu \boldsymbol \Delta = \mathbf 0$
or, equivalently, as
\begin{equation}
\mathbf \Delta = - \mathbf Q \mathbf w \mathbf w^H ( \mathbf w
\mathbf w^H+  \mu \mathbf I)^{-1}. \label{zerograd2}
\end{equation}
Moreover, using the matrix inversion lemma, the expression
\eqref{zerograd2} can be simplified as
\begin{equation}
\mathbf \Delta = - \frac{\mathbf Q \mathbf w \mathbf w^H}
{\|\mathbf w\|^2 + \mu}. \label{Delta}
\end{equation}
The Lagrange multiplier $\mu$ can be determined based on the
conditions \eqref{KKT1}--\eqref{KKT3}. For this goal, we find a
simpler expression for the norm of the matrix $\mathbf Q \mathbf w
\mathbf w^H$ as follows
\begin{eqnarray}
\| \mathbf Q \mathbf w \mathbf w^H \|^2 \!\!& = &\!\! {\rm tr} \{
\mathbf Q \mathbf w \mathbf w^H \mathbf w \mathbf w^H
\mathbf Q^H \} \nonumber \\
\!\!& = &\!\! {\rm tr} \{ \mathbf Q \mathbf w  \mathbf w^H
\mathbf Q^H \} \cdot \mathbf w^H \mathbf w  \nonumber \\
\!\!&=&\!\! \| \mathbf Q \mathbf w \|^2 \| \mathbf w\|^2
\label{proofEq}
\end{eqnarray}
Using \eqref{proofEq}, it can be obtained that
\begin{equation}
\delta (\mu) \triangleq \| \mathbf \Delta\| =  \frac{\|
\mathbf Q \mathbf w\| \| \mathbf w
\|} {\|\mathbf w\|^2 + \mu}. \label{norm}
\end{equation}
{where the new function $\delta (\mu)$ is defined for notation
simplicity. It is easy to verify that $\delta (\mu)$ is a strictly
decreasing function with respect to $\mu \geq 0$. Consequently,
for any arbitrary $\mu \geq 0$, it is true that $\delta (\mu) \leq
\delta (0) = \|\mathbf Q \mathbf w\|/\| \mathbf w\|$. Depending on
whether $\delta (0)$ is less than or equal to $\eta$ or not, the
following two cases are possible. If $\delta (0) \leq \eta$, then
$\mu$ and $\boldsymbol \Delta$ can be found as $\mu=0$ and
$\boldsymbol \Delta = -\mathbf Q \mathbf w \mathbf w^H / \|w
\|^2$, which is obtained by simply substituting $\mu=0$ in
\eqref{Delta}. In this case, the KKT conditions
\eqref{KKT1}--\eqref{KKT3} are obviously satisfied. In the other
case, when $\delta (0)  > \eta$, the above obtained $\mathbf
\Delta$ for $\mu=0$ does not satisfy the condition \eqref{KKT1}
because $\| \mathbf \Delta\| = \delta (0) > \eta$. Since, $\delta
(\mu)$ is a strictly decreasing function with respect to $\mu \geq
0$, for satisfying \eqref{KKT1}, the value of $\mu$ must be
strictly larger than zero and as a result the condition
\eqref{KKT2} implies that $\| \mathbf \Delta \| = \eta$. Note that
if $\mu >0$ and $\| \mathbf \Delta\|$ obtained by substituting
such $\mu$ in \eqref{Delta} is equal to $\eta$, then the KKT
conditions \eqref{KKT1}--\eqref{KKT3} are all satisfied. Thus, we
need to find the value of $\mu$ such that the corresponding $\|
\mathbf \Delta \|$ is equal to $\eta$. By equating $\delta(\mu)$
to $\eta$, it can be resulted that $\mu_0 = \| \mathbf w \|/ \eta
\cdot (\| \mathbf Q \mathbf w \| - \eta \| \mathbf w \|)$.
Considering the above two cases together, the optimal $\mathbf
\Delta$ can be expressed as
\begin{eqnarray}
\mathbf \Delta =\left\{
\begin{array}{ll}
- \eta \frac{ \mathbf Q \mathbf w \mathbf
w^H}{\| \mathbf Q  \mathbf w \| \|\mathbf w \|}  , & \quad
\|\mathbf Q \mathbf w\| \geq  \eta  \| \mathbf w \|  \\
- \frac{ \mathbf Q \mathbf w \mathbf
w^H}{ \|\mathbf w \|^2} , & \quad\  \mbox{otherwise.} \label{DifferentCases}
\end{array}
\right.
\end{eqnarray}
Finally, substituting \eqref{DifferentCases} in the objective
function of the problem \eqref{subsub1}, the worst-case signal
power for a fixed beamforming vector $\mathbf w$ can be found as
shown in \eqref{Lemmaworstcasesol}.} $\hfill\blacksquare$

\subsection{Proof of Lemma 2}
Let $(\mathbf w_{\rm opt},\alpha_{\rm opt})$ denote the optimal
solution of the problem  \eqref{main_problem_alpha}. Let us define
the following auxiliary optimization problem based on the problem
\begin{eqnarray}
\min_{\mathbf w}& &\, \mathbf w^H (\hat{\mathbf {R}} +
\gamma \mathbf  I) \mathbf w \quad \nonumber  \\
{\rm s.t.}  && \mathbf w^H \mathbf Q^H \mathbf Q \mathbf w
= \alpha_{\rm opt}  \nonumber \\
&&\mathbf w^H  \mathbf w \leq \frac{(\sqrt{\alpha_{\rm
opt}}-1)^2}{\eta^2}. \label{main_problem_alpha2}
\end{eqnarray}
It can be seen that if $\mathbf w$ is a feasible point of the
problem  \eqref{main_problem_alpha2}, then the pair $(\mathbf
w,\alpha_{\rm opt})$ is also a feasible point of the problem
\eqref{main_problem_alpha} which implies that the optimal value of
the problem \eqref{main_problem_alpha2} is greater than or equal
to that of \eqref{main_problem_alpha}. However, since $\mathbf
w_{\rm opt}$ is a feasible point of the problem
\eqref{main_problem_alpha2} and the value of the objective
function at this feasible point is equal to the optimal value of
the problem \eqref{main_problem_alpha}, i.e., it is equivalent to
$\mathbf w_{\rm opt}^H (\hat{\mathbf {R}} + \gamma \mathbf  I)
\mathbf w_{\rm opt}$, it can be concluded that both of the
optimization problems \eqref{main_problem_alpha} and
\eqref{main_problem_alpha2} have the same optimal value. Let us
define another auxiliary optimization problem based on the problem
\eqref{main_problem_alpha2} as
\begin{eqnarray}
g \triangleq &\min\limits_{\mathbf w}& \, \mathbf w^H
(\hat{\mathbf {R}} + \gamma \mathbf I) \mathbf w \quad  \nonumber\\
&{\rm s.t.}&  \mathbf w^H  \mathbf Q^H \mathbf Q \mathbf w =
\alpha_{opt} \label{auxiliary_problem_org3}
\end{eqnarray}
which is obtained from \eqref{main_problem_alpha2} by dropping the
last constraint of \eqref{main_problem_alpha2}. The feasible set
of the optimization problem \eqref{main_problem_alpha2} is a
subset of the feasible set of the optimization problem
\eqref{auxiliary_problem_org3}. As a result, the optimal value $g$
of the problem \eqref{auxiliary_problem_org3} is smaller than or
equal to the optimal value of the problem
\eqref{main_problem_alpha2}, and thus also, the optimal value of
the problem \eqref{main_problem_alpha}. Using the maxmin theorem
\cite{Haykin}, it is easy to verify that $g= \alpha_{opt} /
\lambda_{\rm max}\left\{( \hat{\mathbf {R}} + \gamma \mathbf
I)^{-1} \mathbf Q^{H} \mathbf Q \right\}$. Since $g$ is smaller
than or equal to the optimal value of the problem
\eqref{main_problem_alpha}, it is upper-bounded by $\mathbf
w_0^H(\hat{\mathbf {R}} + \gamma \mathbf I) \mathbf w_0$, where
$\mathbf w_0$ is an arbitrary feasible point of
\eqref{main_problem_alpha}. The latter implies that $ \alpha_{opt}
\leq  \lambda_{\rm max}\left\{( \hat{\mathbf {R}} + \gamma \mathbf
I)^{-1} \mathbf Q^{H} \mathbf Q \right\} \cdot \mathbf w_0^H
(\hat{\mathbf {R}} + \gamma \mathbf I) \mathbf w_0$.
$\hfill\blacksquare$

\subsection{Proof of Lemma 3}
It is easy to verify that the dual problem for both optimization
problems of $h(\alpha)$ and $k(\alpha)$ is the same and it can be
expressed for fixed $\alpha$ as
\begin{eqnarray}
\max \limits_{\tau, \psi}  && \tau \alpha -\psi ( \sqrt{\alpha} -
1)^2/\eta^2 \nonumber \\
{\rm s.t.}  && (\hat{\mathbf {R}} + \gamma \mathbf I) -\tau \cdot
\mathbf Q^H \mathbf Q + \psi \mathbf I \succeq \mathbf 0 \nonumber \\
&& \psi \geq 0 \label{dualproblem}
\end{eqnarray}
where $\tau$ and $\psi$ are the Lagrange multipliers. Based on the
dual problem \eqref{dualproblem}, a new optimal value function is
defined as
\begin{eqnarray}
d(\alpha) \!\!&\triangleq &\!\! \Big \{ \max \limits_{\tau, \psi}
\tau \alpha -\psi \frac{(\sqrt{\alpha}-1)^2}{\eta^2} \ | \
(\hat{\mathbf {R}} + \gamma \mathbf I) \nonumber \\
\!\!&-&\!\! \tau \mathbf Q^H \mathbf Q + \psi \mathbf I \succeq
\mathbf 0, \ \psi \geq 0 \Big \}.
\end{eqnarray}

The optimization problem of $k(\alpha)$ is a convex SDP problem
which satisfies the Slater's conditions as $\tau=0$ and $\psi=1$
is a strictly feasible point for its dual problem
\eqref{dualproblem}. Thus, the duality gap between the
optimization problem of $h(\alpha)$, i.e., the problem
\eqref{hdef1} and its dual problem \eqref{dualproblem} is zero. It
implies that
\begin{equation}
k(\alpha)=d(\alpha), \quad \alpha \in [\theta_1,\theta_2].
\label{equation1}
\end{equation}

On the other hand, the optimization problem of $h(\alpha)$ is
specifically a quadratically constrained quadratic programming
(QCQP) problem with only two constraints. It has been recently
shown that the duality gap between a QCQP {in complex variables}
with two constraints and its dual problem is zero
{\cite{AmirEldar},} \cite{Daniel}. Based on the latter fact, it
can be obtained that
\begin{equation}
h(\alpha)=d(\alpha), \quad \alpha \in [\theta_1,\theta_2].
\label{equation2}
\end{equation}
Using \eqref{equation1} and \eqref{equation2}, it can be concluded
that the functions $h(\alpha)$ and $k(\alpha)$ are equivalent.

Let $\mathbf w_{\alpha}$ denote the optimal solution of the
optimization problem of $h(\alpha)$ which implies that $h(\alpha)
= \mathbf w_{\alpha}^H (\hat{\mathbf {R}} \!+\! \gamma \mathbf I)
\mathbf w_{\alpha}$. It is then trivial to verify that $\mathbf
W_{\alpha} \triangleq \mathbf w_{\alpha} \mathbf w_{\alpha}^H$ is
a feasible point of the optimization problem of $k(\alpha)$ and
${\rm tr}\left\{ (\hat{\mathbf {R}} + \gamma \mathbf I)  \mathbf
W_{\alpha} \right\}=h(\alpha)$. Using the fact that $h(\alpha)$
and $k(\alpha)$ are equivalent, it can be concluded that ${\rm
tr}\left\{ (\hat{\mathbf {R}} + \gamma \mathbf I) \mathbf
W_{\alpha} \right\}=k(\alpha)$ which implies that $\mathbf
W_{\alpha}$ is the optimal solution of the optimization problem of
$k(\alpha)$. The latter means that, for a fixed value of $\alpha$,
the optimization problem of $k(\alpha)$ has always a rank-one
solution. A method for extracting a rank-one solution from a
general-rank solution of QCQP is explained, for example, in
\cite{Daniel}. It is trivial to see that the scaled dominant
eigenvector of such a rank-one solution of the optimization
problem of $k(\alpha)$ is the optimal solution of the optimization
problem of $h(\alpha)$. $\hfill\blacksquare$

\subsection{Proof of Lemma 4}
First, we prove that $l(\alpha,\alpha_c)$ is a convex function
with respect to $\alpha$. For this goal, let $\mathbf
W_{\alpha_1}$ and $\mathbf W_{\alpha_2}$ denote the optimal
solution of the optimization problems of $l(\alpha_1,\alpha_c)$
and $l(\alpha_2,\alpha_c)$, respectively, i.e.,
$l(\alpha_1,\alpha_c)={\rm tr}\left\{ (\hat{\mathbf {R}} + \gamma
\mathbf I)  \mathbf W_{\alpha_1} \right\}$ and
$l(\alpha_2,\alpha_c)={\rm tr}\left\{ (\hat{\mathbf {R}} + \gamma
\mathbf I)  \mathbf W_{\alpha_2} \right\}$,  where $\alpha_1$ and
$\alpha_2$ are any two arbitrary points in the interval
$[\theta_1,\theta_2]$. It is trivial to verify that $\theta
\mathbf W_{\alpha_1} + (1-\theta) \mathbf W_{\alpha_2}$ is a
feasible point of the corresponding optimization problem of
$l(\theta \alpha_1 + (1-\theta) \alpha_2,\alpha_c)$ (see the
definition \eqref{Ldef}). Therefore,
\begin{eqnarray}
l(\theta  \alpha_1 \!\!\!\!&+&\!\!\!\!\! (1\!-\!\theta)
\alpha_2,\alpha_c) \nonumber \\ &&\leq {\rm tr} \big\{
(\hat{\mathbf {R}} \!+\! \gamma \mathbf I) (\theta \mathbf
W_{\alpha_1}\!\!+\! (1\!-\!\theta) \mathbf W_{\alpha_2}) \big\}
\nonumber \nonumber \\&&= \theta {\rm tr}\big\{ (\hat{\mathbf {R}}
+ \gamma \mathbf I)  \mathbf W_{\alpha_1} \big\} \nonumber \\&&\ \
+ (1-\theta) {\rm tr}\big\{ (\hat{\mathbf {R}} + \gamma \mathbf I)
\mathbf W_{\alpha_2} \big\} \nonumber \\&&= \theta
l(\alpha_1,\alpha_c) + (1-\theta) l(\alpha_2,\alpha_c)
\end{eqnarray}
which proves that $l(\alpha,\alpha_c)$ is a convex function with
respect to $\alpha$.

In order to show that $l(\alpha,\alpha_c)$ is greater than or
equal to $k(\alpha)$, it suffices to show that the feasible set of
the optimization problem of $l(\alpha,\alpha_c)$ is a subset of
the feasible set of the optimization problem of $k(\alpha)$. Let
$\mathbf W_1$ denote a feasible point of the optimization problem
of $l(\alpha,\alpha_c)$, it is easy to verify that $\mathbf W_1$
is also a feasible point of the optimization problem of
$k(\alpha)$ if the inequality $\sqrt{\alpha} \leq \sqrt{\alpha_c}
+ \frac{\alpha - \alpha_c}{2 \sqrt{\alpha_c}}$ holds. This
inequality can be rearranged as
\begin{equation}
(\sqrt{\alpha}-\sqrt{\alpha_c})^2 \geq 0
\end{equation}
and it is valid for any arbitrary $\alpha$. Therefore, $\mathbf
W_1$ is also a feasible point of the optimization problem of
$k(\alpha)$ which implies that $l(\alpha,\alpha_c) \geq
k(\alpha)$.

In order to show that the right and left derivatives are equal, we
use the result of \cite[Theorem 10]{Alexander} which gives
expressions for the directional derivatives of a parametric SDP.
Specifically, in \cite[Theorem 10]{Alexander} the directional
derivatives for the following optimal value function
\begin{equation}
\psi(\mathbf u) \triangleq \{\min\limits_{\mathbf y} f(\mathbf
y,\mathbf u)\ | \ \mathbf G(\mathbf y,\mathbf u) \preceq \mathbf
0_{n \times n} \} \label{paraprob}
\end{equation}
are derived, where $f(\mathbf y,\mathbf u)$ and $\mathbf G(\mathbf
y,\mathbf u)$ are a scaler and an $n \times n$ matrix,
respectively, $\mathbf y \in {\cal R}^m$ is the optimization
variables and $\mathbf u \in {\cal R}^k$ is the optimization
parameters. Let $\mathbf u_c$ be an arbitrary fixed point. If the
optimization problem of $\psi(\mathbf u_c)$ poses certain
properties, then according to \cite[Theorem 10]{Alexander} it is
directionally differentiable at $\mathbf u_c$. These properties
are (i) the functions $f(\mathbf y,\mathbf u)$ and $\mathbf
G(\mathbf y,\mathbf u)$ are continuously differentiable, (ii) the
optimization problem of $\psi(\mathbf u_c)$ is convex, (iii) the
set of optimal solutions of the optimization problem of
$\psi(\mathbf u_c)$ denoted as $\cal M$ is nonempty and bounded,
(iv) the Slater condition for the optimization problem of
$\psi(\mathbf u_c)$  holds true, and (v) the {\it inf-compatness}
condition is satisfied. Here inf-compatness condition refers to
the condition of the existence of  $\alpha > \psi(\mathbf u_c)$
and  a compact set $S \subset {\cal R}^m$ such that $ \{ \mathbf y
| f(\mathbf y,\mathbf u) \leq \alpha, \mathbf G(\mathbf y,\mathbf
u) \preceq \mathbf 0\}  \subset S $ for all $\mathbf u$ in a
neighborhood of $\mathbf u_c$. If for all $\mathbf u$ the
optimization problem of $\psi(\mathbf u)$ is convex and the set of
optimal solutions of $\psi(\mathbf u)$ is non-empty and bounded,
then the inf-compactness conditions holds automatically.

The directional derivative of $\psi(\mathbf u)$ at $\mathbf u_c$
in a direction $\mathbf d \in {\cal R}^k$ is given by
\begin{equation}
\psi^\prime(\mathbf u_c, \mathbf d)= \min \limits_{\mathbf y \in
{\cal M}} \max\limits_{\boldsymbol \Omega \in {\cal Z}} \mathbf
d^T \nabla_{\mathbf u} L(\mathbf y,\boldsymbol \Omega, \mathbf
u_c),
\end{equation}
where $\cal{Z}$ is the set of optimal solutions of the dual
problem of the optimization problem of $\psi(\mathbf u_c)$ and
$L(\mathbf y,\Omega, \mathbf u)$ denotes the Lagrangian defined as
\begin{equation}
L(\mathbf y,\Omega, \mathbf u) \triangleq f(\mathbf y,\mathbf u) +
{\rm tr}\left(\boldsymbol \Omega \cdot \mathbf G(\mathbf y,
\mathbf u)\right)
\end{equation}
where $\boldsymbol \Omega$ denotes the Lagrange multiplier matrix.

Let us look again to the definitions of the optimal value
functions $k(\alpha)$ and $l(\alpha,\alpha_c)$ \eqref{kalpha} and
\eqref{Ldef}, respectively, and define the following block
diagonal matrix
\begin{eqnarray}
&&\!\!\!\!\!\!\!\mathbf G_1 (\mathbf W,\alpha)= \\ && \!\!\!\!\!\!\!
\begin{pmatrix}
-\mathbf W \!\!\!\!&\!\!\!\! \mathbf 0 \!\!\!\!\!\!\!\!\!\!\!&\!\!
\!\!\!\!\!\!\!\!\!\!\! \mathbf 0 \!\!\!\!\!\!\!\!\!&\!\! \!\!\!\!
\! \! \! \mathbf 0 \\
\mathbf 0 \!\!\!\!&\!\!\!\! {\eta^2} {\rm tr}\{ \mathbf W\}\!-\!{(
\sqrt{\alpha}\!-\!1)^2} \!\!\!\!\!\!\!\!\!\!\!&\!\!\!\!\!\!\!\!\!
\!\!\!\! 0\!\!\!\!\!\!\!\!\!&\!\!\!\!\!\!\!\!\! 0 \\
\mathbf 0  \!\!\!\!&\!\!\!\!\!\!\! 0  \!\!\!\!\!\!\!\!\!\!\!&\!\!
\!\!\!\!\!\!\!\!\!\!\! {\rm tr}\{  \mathbf Q^H \mathbf Q \mathbf W
\} - \alpha \!\!\!\!\!\!\!\!\!&\!\!\!\!\!\!\!\!\! 0  \\
\mathbf 0 \!\!\!\!&\!\!\!\!\!\!\! 0
\!\!\!\!\!\!\!\!\!\!\!&\!\!\!\!\!\!\!\!\!\!\!\!\! 0
\!\!\!\!\!\!\!\!\!&\!\!\!\!\!\!\!\!\!   \alpha - {\rm tr}\{
\mathbf Q^H \mathbf Q \mathbf W\} \nonumber
\end{pmatrix}
\end{eqnarray}
as well as another block diagonal matrix denoted as $\mathbf G_2
(\mathbf W,\alpha)$ which has exactly same structure as the matrix
$\mathbf G_1 (\mathbf W,\alpha)$ with only difference that the
element ${\eta^2} {\rm tr}\{ \mathbf W\}-{(\sqrt{\alpha}-1)^2}$ in
$\mathbf G_1 (\mathbf W,\alpha)$ is replaced by ${\eta^2} \cdot
{\rm tr} \{ \mathbf W\! \} +{(\sqrt{\alpha_c}-1) + \alpha \left(
{1} / {\sqrt{\alpha_c}}-1\right)}$ in $\mathbf G_2 (\mathbf
W,\alpha)$.
%\begin{eqnarray}
%&&\!\!\!\!\!\!\!\!\!\!\!\mathbf G_2 (\mathbf W,\alpha)= \nonumber \\ &&
%\!\!\!\!\!\!\!\!\begin{pmatrix}
%   -\mathbf W \!\!\!\!&\!\!\!\!  \mathbf 0 \!\!\!\!\!\!\!\!\!\!\!&\!\!\!\!\!\!\!\!\!\!\!\!\! \mathbf 0 \!\!\!\!\!\!\!\!\!&\!\!\!\!\!\!\!\!\! \mathbf 0 \\
%    \mathbf 0 \!\!\!\!&\!\!\!\! {\eta^2} {\rm tr} \{ \mathbf W\! \} \!+\!\!(\sqrt{\alpha_c}\!-\!1) \!\!+\!\!
%\alpha(\frac{1}{\sqrt{\alpha_c}}\!-\!1)\! \!\!\!\!\!\!\!\!\!\!\!&\!\!\!\!\!\!\!\!\!\!\!\!\! 0\!\!\!\!\!\!\!\!\!&\!\!\!\!\!\!\!\!\! 0 \\
%   \mathbf 0  \!\!\!\!&\!\!\!\!\!\!\! 0  \!\!\!\!\!\!\!\!\!\!\!&\!\!\!\!\!\!\!\!\!\!\!\!\! {\rm tr}\{  \mathbf Q^H \mathbf Q \mathbf W\} - \alpha \!\!\!\!\!\!\!\!\!&\!\!\!\!\!\!\!\!\! 0  \\
%   \mathbf 0 \!\!\!\!&\!\!\!\!\!\!\! 0 \!\!\!\!\!\!\!\!\!\!\!&\!\!\!\!\!\!\!\!\!\!\!\!\! 0 \!\!\!\!\!\!\!\!\!&\!\!\!\!\!\!\!\!\!   \alpha - {\rm tr}\{  \mathbf Q^H \mathbf Q \mathbf W\}
%\end{pmatrix} \nonumber
%\end{eqnarray}
%\begin{equation}
%\mathbf G_2 (\mathbf W,\alpha)=
%\begin{pmatrix}
%   -\mathbf W &  \mathbf 0_{1 \times M} & \mathbf 0_{M \times 1} & \mathbf 0_{M \times 1} \\
%    \mathbf 0_{M \times 1} & {\eta^2} \cdot {\rm tr} \{ \mathbf W\! \} \!+{\!(\sqrt{\alpha_c}\!-\!1) +
%\alpha \left({1}/{\sqrt{\alpha_c}}\!-\!1\!\right)} \leq \!0 & 0& 0 \\
%   \mathbf 0_{1 \times M}  & 0  & {\rm tr}\{  \mathbf Q^H \mathbf Q \mathbf W\} - \alpha & 0  \\
%   \mathbf 0_{1 \times M} & 0 & 0 &   \alpha - {\rm tr}\{\mathbf Q^H \mathbf Q \mathbf W\}
%\end{pmatrix}
%\end{equation}
Then the optimal value functions $k(\alpha)$ and $l(\alpha,
\alpha_c)$ can be equivalently recast as
\begin{eqnarray}
k(\alpha) \!\!\!&=&\!\!\! \left \{ \min\limits_{\mathbf W} {\rm
tr}\left\{ (\hat{\mathbf {R}} + \gamma \mathbf I) \cdot \mathbf W
\right\} \ |  \mathbf G_1 (\mathbf W,\alpha) \preceq \mathbf 0
\right \},\nonumber \\&& \quad \theta_1 \leq \alpha \leq \theta_2
\end{eqnarray}
and
\begin{eqnarray}
l(\alpha,\alpha_c) \!\!\!&=&\!\!\! \left\{ \min\limits_{\mathbf W}
{\rm tr}\left\{ (\hat{\mathbf {R}} + \gamma \mathbf I) \cdot
\mathbf W \right\} \ |  \mathbf G_2 (\mathbf W,\alpha) \preceq
\mathbf 0   \right \},\nonumber \\&&  \quad \theta_1 \leq \alpha
\leq \theta_2.
\end{eqnarray}

It is trivial to verify that the optimization problems of
$k(\alpha_c)$ and $l(\alpha_c,\alpha_c)$ can be expressed as
\begin{eqnarray}
&\min\limits_{\mathbf W}& {\rm tr}\{ (\hat{\mathbf {R}} + \gamma
\mathbf I) \cdot \mathbf W \} \nonumber  \\
&{\rm s.t.}& \ {\rm tr}\{  \mathbf Q^H \mathbf Q \mathbf W\} =
\alpha_c \nonumber \\
&& \ {\rm tr}\{ \mathbf W\} \!\leq\!
\frac{(\sqrt{\alpha_c}\!-\!1)^2}{\eta^2} \nonumber \\ && \ \mathbf
W \succeq \mathbf 0 . \label{unpertprob}
\end{eqnarray}
The problem \eqref{unpertprob} is convex and its solution set is
non-empty and bounded. Indeed, let $\mathbf W_1$ and $\mathbf W_2$
denote two optimal solutions of the problem above. The Euclidean
distance between $\mathbf W_1$ and $\mathbf W_2$ can be expressed
as
\begin{eqnarray}
d(\mathbf W_1,\mathbf W_2) \!\!\!&=&\!\!\! \|\mathbf W_1 -
\mathbf W_2 \| \nonumber \\
\!\!\!&=&\!\!\! \sqrt{{\rm tr}\{\mathbf W_1^2\} + {\rm tr}\{
\mathbf W_2^2\} -2  {\rm tr}\{\mathbf W_1 \mathbf W_2\}}
\nonumber \\
\!\!\!&\leq&\!\!\! \sqrt{2 \frac{(\sqrt{\alpha_c}\!-\!1)^4}
{\eta^4}}
\end{eqnarray}
where the last line is due to the fact that the matrix product
$\mathbf W_1 \mathbf W_2$ is positive semi-definite and,
therefore, ${\rm tr}\{\mathbf W_1 \mathbf W_2\} \geq 0$, and also
the fact that for any arbitrary positive semi-definite matrix
${\rm tr} \{\mathbf A^2\} \leq {\rm tr} \{\mathbf A\}^2$. From the
equation above, it can seen that the distance between any two
arbitrary optimal solutions of \eqref{unpertprob} is finite and,
therefore, the solution set is bounded. As it was mentioned in the
proof of Lemma~2, the optimization problem \eqref{unpertprob}
satisfies the strong duality. In a similar way, it can be shown
that the inf-compactness condition is satisfied by verifying that
the optimization problems of $k(\alpha)$ and $l(\alpha,\alpha_c)$
are convex and their corresponding solution sets are bounded for
any $\alpha$. Therefore, both of the optimal value functions
$k(\alpha)$ and $l(\alpha,\alpha_c)$ are directionally
differentiable at $\alpha_c$.

Using the result of \cite[Theorem 10]{Alexander}, the directional
derivatives of $k(\alpha)$ and $l(\alpha,\alpha_c)$ can be
respectively computed as
\begin{equation}
k^{\prime}(\alpha,d)=\min \limits_{\mathbf W \in {\cal M}}
\max\limits_{\boldsymbol \Omega \in {\cal Z}} d  \left({\rm
tr}\big \{\boldsymbol \Omega \cdot \frac{d}{d \alpha}\mathbf G_1
(\mathbf W,\alpha)\big|_{\alpha=\alpha_c}\big\}  \right)
\end{equation}
and
\begin{equation}
l^{\prime}(\alpha,\alpha_c,d)=\min \limits_{\mathbf W \in {\cal
M}} \max\limits_{\boldsymbol \Omega \in {\cal Z}} d \left ({\rm
tr}\big\{\boldsymbol \Omega \cdot \frac{d}{d \alpha}\mathbf G_2
(\mathbf W,\alpha)\big|_{\alpha=\alpha_c}\big\}\right)
\end{equation}
where ${\cal M}$ and ${\cal Z}$ denote the optimal solution sets
of the optimization problem of \eqref{unpertprob} and its dual
problem, respectively. Using the definitions of  $\mathbf G_1
(\mathbf W,\alpha)$ and $\mathbf G_2 (\mathbf W,\alpha)$, it can
be seen that the terms ${d} \mathbf G_1 (\mathbf W,\alpha) / {d
\alpha}$ and ${d}\mathbf G_1 (\mathbf W,\alpha)/{d \alpha}$ are
equal at $\alpha=\alpha_c$ and, therefore, the directional
derivatives are equivalent. The latter implies that the left and
right derivatives of  $k(\alpha)$ and $l(\alpha,\alpha_c)$ are
equal at $\alpha=\alpha_c$. $\hfill\blacksquare$

\subsection{Proof of Lemma 5}
i) As it has been explained, the optimization problem in
Algorithm~1 at iteration $i,\ i\geq 2$ is obtained by linearizing
$\sqrt{\alpha}$ at $\alpha_{opt,i-1}$. Since $\mathbf W_{opt,i-1}$
and $\alpha_{opt,i-1}$ are feasible for the optimization problem
at iteration $i$, it can be straightforwardly concluded that the
optimal value of the objective at iteration $i$ is less than or
equal to the optimal value at the previous iteration, i.e., ${\rm
tr} \left\{ (\hat{\mathbf {R}} + \gamma \mathbf I) \mathbf
W_{opt,i}\right\} \leq {\rm tr} \left\{ (\hat{\mathbf {R}} +
\gamma \mathbf I) \mathbf W_{opt,i-1} \right\}$ which completes
the proof.

ii) Since the sequence of the optimal values, i.e., ${\rm tr}
\left\{ (\hat{\mathbf {R}} + \gamma \mathbf I) \mathbf W_{opt,i}
\right\}, i \geq 1$ is non-increasing and bounded from below
(every optimal value is non-negative), the sequence of the optimal
values converges.

iii) The proof follows straightforwardly from Proposition 3.2 of
\cite[Section 3]{Amir}. $\hfill\blacksquare$

\end{document}